\documentclass[preprint2,longabstract]{aastex}

\slugcomment{}

\shorttitle{Discovery of Ultra-Compact Dwarfs in Virgo}
\shortauthors{J. B. Jones, M. J. Drinkwater, R. Jurek et al.}

\begin{document}


\title{Discovery of Ultra-Compact Dwarf Galaxies in the Virgo Cluster}


\author{J. B. Jones}
\affil{Astronomy Unit, School of Mathematical Sciences, 
                Queen Mary University of London, Mile~End~Road, 
                London, E1~4NS, U.K. }

\author{M. J. Drinkwater and R. Jurek}
\affil{Department of Physics, University of Queensland, 
                Queensland 4072, Australia}
\author{S. Phillipps}
\affil{Astrophysics Group, Department of Physics, 
                University of Bristol, Tyndall Avenue, Bristol, 
                BS8 1TL, U.K.}
\author{M. D. Gregg\altaffilmark{1}}
\affil{Department of Physics, University of California, Davis, 
                CA 95616, U.S.A.}
\author{K. Bekki and W. J. Couch}
\affil{School of Physics, University of New South Wales, 
                Sydney 2052, Australia}
\author{A. Karick}
\affil{Institute for Geophysics and Planetary Physics, 
                Lawrence Livermore National Laboratory, L-413, 
                Livermore, CA 94550, U.S.A.}
\author{Q. A. Parker\altaffilmark{2}}
\affil{Department of Physics, Macquarie University,
                Sydney, New South Wales 2109, Australia}
\and

\author{R. M. Smith}
\affil{School of Physics and Astronomy, University of Wales Cardiff, 
                P.O. Box 913, Cardiff, CF12~3YB, U.K.}


\altaffiltext{1}{Institute for Geophysics and Planetary Physics, 
                Lawrence Livermore National Laboratory, L-413, 
                Livermore, CA 94550, U.S.A.}
\altaffiltext{2}{Anglo-Australian Observatory, P. O. Box 296, Epping,
                NSW 1710, Australia.}


\begin{abstract}

We have discovered nine ultra-compact dwarf galaxies (UCDs) in the
Virgo Cluster, extending samples of these objects 
outside the Fornax Cluster.  Using the 2dF multi-fiber
spectrograph on the Anglo-Australian Telescope, the new Virgo members
were found among 1500 color-selected, star-like targets with $16.0 <
b_j < 20.2$ in a two-degree diameter field centered on M87 (NGC4486). 
The newly-found UCDs are comparable to the UCDs in the Fornax Cluster, 
with sizes $\lesssim 100$~pc, $-12.9 < M_B < -10.7$, 
and exhibiting red, absorption-line 
spectra, indicative of an older stellar population.  The properties of
these objects remain consistent with the tidal threshing model for the
origin of UCDs from the surviving nuclei of nucleated dwarf
ellipticals disrupted in the cluster core, but can also be explained 
as objects that were formed by mergers of star clusters created in 
galaxy interactions.  The discovery that UCDs
exist in Virgo shows that this galaxy type is probably a ubiquitous
phenomenon in clusters of galaxies; coupled with their possible origin
by tidal threshing, the UCD population is a potential indicator and
probe of the formation history of a given cluster.

We also describe one additional bright UCD with $M_B =
-12.0$ in the core of the Fornax Cluster.  We find no further UCDs in
our Fornax Cluster Spectroscopic Survey down to $b_j=19.5$ in two
additional 2dF fields extending as far as $3\degr$ from the center
of the cluster.  All six Fornax bright UCDs identified with 2dF lie
within $0\fdg5$ (projected distance of 170~kpc) of the central
elliptical galaxy NGC1399.

\end{abstract}





\section{Introduction}
\label{sec_intro}

Compact dwarf galaxies are an established constituent of the
population of lower luminosity galaxies, alongside the more numerous
dwarf irregular, dwarf elliptical and dwarf spheroidal galaxies. Of
compact dwarfs, early-type compact ellipticals like M32, showing
early-type galaxy spectra, are rare and only very few examples have been
found (Faber 1973, Nieto \& Prugniel 1987).  Blue compact dwarfs show
active star formation superimposed on an older low surface brightness
component, and have a broad range in sizes, including a very small
proportion with effective radii as small as $300\;$pc (e.g.\ Doublier,
Caulet \& Comte 1999;  Cair\'{o}s et al.\ 2001;  Gil de Paz, Madore \&
Pevunova 2003).  The extremely compact star-forming object POX186 has
been described by, among others, Kunth et al.\ (1988), Doublier et
al.\  (2000) and Corbin \& Vacca (2002).

However, a new class of compact galaxy, more than an order of
magnitude smaller in physical size than conventional compact dwarfs,
has recently been discovered in the Fornax Cluster. Hilker et
al.\  (1999) found two very compact objects in a spectroscopic survey of
the cluster with a fiber-fed spectrograph on the 2.5-metre du Pont
Telescope. Meanwhile, Drinkwater et al.\ (2000a; see also Phillipps et
al.\  2001; Drinkwater et al.\ 2002) identified five objects that had
velocities of cluster galaxies and which were either unresolved or
marginally resolved in ground-based imaging during the Fornax Cluster
Spectroscopic Survey (FCSS; Drinkwater et al.\ 2000b), including the 
two Hilker et al.\ objects.
The five compact objects had blue absolute magnitudes $-13
\leq M_B \leq -11$, and sizes $\ll 100$~pc: they were therefore named
``ultra-compact dwarfs" (UCDs).  

Possible explanations for these ultra-compact dwarfs include unusually
luminous globular clusters (e.g.\ Hilker et al.\ 1999; Drinkwater et
al.\  2000a; Phillipps et al.\ 2001; Mieske, Hilker \& Infante 2002),
extremely luminous star clusters formed in galaxy interactions
(Fellhauer \& Kroupa 2002), low-luminosity analogues to M32
(Drinkwater et al.\ 2000a), and the nuclei of very low surface
brightness galaxies (Phillipps et al.\ 2001).  Fellhauer \& Kroupa 
used numerical simulations to model the formation of star clusters 
that would evolve into objects similar to UCDs. 
Some authors have argued
that highly compact galaxies might have been formed in the early
Universe (Blanchard, Valls-Gabaud \& Mamon 1992; Tegmark et al.\ 1997).

Drinkwater et al.\ (2000a) and Phillipps et al.\ (2001) argued however
that the UCD luminosities, their distribution within the Fornax
Cluster, and that one object was marginally resolved in
ground-based imaging, implied that they were most likely to be the
nuclei of tidally-stripped nucleated dwarf ellipticals (dE,Ns).
Bekki, Couch, \& Drinkwater (2001), building on the simulations of
Bassino, Muzzio, \& Rabolli (1994), modelled the tidal destruction of
dE,Ns by nearby giant galaxies, calling the process tidal threshing,
and predicted that the remnants will have properties similar to UCDs.

More recently, Drinkwater et al.\ (2003) 
and Gregg et al.\ (2003) 
have presented results for the five bright Fornax Cluster UCDs using STIS
imaging from the Hubble Space Telescope and high-resolution
spectroscopy from the European Southern Observatory Very Large
Telescope and the Keck II Telescope.  All five UCDs were spatially
resolved and have effective radii between 10 and 22~pc, while one was
also surrounded by a lower surface brightness compact halo.  
These are larger than conventional
globular clusters. The velocity dispersions of their stellar
populations were found to be in the range 24 to 37~km$\,$s$^{-1}$,
larger than Galactic globular clusters but just overlapping the most
luminous globular clusters in M31.  They lie away from the globular
cluster sequence in the luminosity -- velocity dispersion plane, but
again close to the nuclei of nucleated dwarf elliptical galaxies.
Drinkwater et al.\ (2003) interpreted these results as supporting the
hypothesis that UCDs are the remnant nuclei of threshed dE,Ns.

Bekki et al.\ (2003) performed detailed numerical simulations of the
tidal disruption of dE,N galaxies in the vicinity of massive galaxies.
They investigated the sensitivity of the disruption process to the
dE,N mass, the dE,N orbital parameters and dE,N dark matter
profile. Under some circumstances the disruption will be incomplete,
leaving a low surface brightness halo around the dE,N nucleus.
Significantly, a cuspy dark matter distribution, such as in a Navarro,
Frenk \& White (1996) profile, can inhibit the threshing process.
Similarly, Kazantzidis, Moore \& Mayer (2003) performed N-body
simulations of the tidal stripping of dE,N galaxies within the core of
a galaxy cluster, including the effects of baryonic matter, to
quantify the stripping efficiency as a function of the concentration
of Navarro-Frenk-White dark matter profiles (see also Moore 2003):
complete dE,N disruption can only occur if the profiles have very low
central concentrations.  Thus if UCDs are indeed produced by the tidal 
threshing of dE,Ns, it follows that UCD observations might be a
powerful test of the profiles of dark matter halos.

Recently, two further surveys of the central region of the Fornax
Cluster to fainter magnitudes than the Phillipps et al.\ (2001) 2dF
study (which extended to $b_j$ = 19.7) have found a large population
of fainter compact objects.  Mieske, Hilker \& Infante (2004) have
found 54 new objects within $20\arcmin$  of NGC1399 to V = 21.0~mag (to
$M_B \simeq -9.8$~mag).  Meanwhile Drinkwater et al.\ (2004, and in 
preparation) have identified a similar number of compact objects to 
B $\simeq 21.5$~mag extending far beyond the normal globular cluster 
population of NGC1399 in radial extent (as far as $0.9^\circ$), 
indicating that the brighter UCDs are accompanied by a much more 
numerous population of fainter counterparts (which will include 
intracluster globular clusters; see Bassino et al. 2003). 

De Propris et al. (2005) compared Hubble Space Telescope imaging of 
the five Drinkwater et al. (2000a) Fornax UCDs with nuclei of dwarf 
eilliptical galaxies, finding some differences in size and surface 
brightness. 
Mieske et al. (2005) have identified candidate UCDs from HST imaging of 
the Abell~1689 cluster. 
Ha\c{s}egan et al. (2005) used HST imaging and Keck telescope spectroscopy 
to identify compact objects 
in individual fields within the Virgo Cluster to investigate the 
relationship between UCDs and the most luminous globular clusters. 

To test whether UCDs exist in other clusters, we undertook 2dF
spectroscopic observations in Virgo.  The central galaxies in Fornax
(NGC1399) and Virgo (M87) have similar luminosities, sizes and
classifications, while the two clusters have similar distributions of
nucleated dwarf ellipticals (dE,Ns) in their central regions
(Binggeli, Sandage \& Tammann 1985; Ferguson 1989; Ferguson \& Sandage
1988).  The galaxy threshing model therefore predicts a population of
UCDs in the vicinity of M87, yet because the Virgo Cluster is several
times less dense than Fornax, one might expect fewer UCDs per volume.
Although the radial distribution of the Virgo dE,Ns is slightly less
concentrated than that for Fornax, there are significantly more dE,Ns
in the same projected area (56 in the region covered by a two-degree
diameter 2dF field centered on M87, compared to 37 for the equivalent
Fornax 2dF field centered on NGC1399).  Naively scaling our UCD numbers
from Fornax, we expected to find $9\pm3$ UCDs in a single 2dF
field centered on the M87 core of the Virgo Cluster.  Bekki et al.\
(2003) provided more precise predictions of a population of UCDs in
Virgo, using detailed N-body simulations, and demonstrated that
identifying UCDs in a new environment, and measuring their detailed
properties, would provide a test of formation models.

Informed by the properties of the Fornax objects, we targeted a
restricted color range of unresolved objects, concentrating within
$0\fdg65$ of M87.  With just 6 hours of 2dF service time we
observed 1501 objects, discovering 9 Virgo UCDs within 14 to 150 kpc
of M87.  We report the discovery of these new UCDs in
Section~\ref{sec_virgo} of this paper (see also Drinkwater et al. 
2004).  We assume a Virgo Cluster
distance modulus of 31.0~mag (see, for example, Ferarese et al.\  2000), 
equivalent to 16~Mpc, and a Fornax
Cluster distance modulus of 31.5~mag, equivalent to 20~Mpc
(Drinkwater, Gregg \& Colless 2001).

Since the original FCSS Fornax discoveries, we have improved the
completeness in the first 2dF field and completed observations of two
more.  We report here, in Section~\ref{sec_fornax}, the discovery of
one new Fornax UCD in FCSS Field 1 centered on NGC1399, and null
results for the other two fields.

\section{A Survey for Ultra-compact Dwarf Galaxies in the Virgo Cluster}
\label{sec_virgo}

\subsection{A targetted 2dF search of the Virgo Cluster M87 core}
\label{sec_virgotargets}

The discovery of UCDs in the Fornax Cluster raises the questions of
how common this type of object is in clusters in general, and whether
the local environment in a cluster affects their number densities or
properties.  To address these issues, we have carried out a search for
similar objects in a single 2dF field in the Virgo Cluster centered on
the giant elliptical galaxy M87.

The five original 2dF Fornax UCDs were found from a nearly-complete sample of
targets defined only by apparent magnitude and position, irrespective
of morphology or color, at the expense of considerable observing time
to obtain spectra for every such object.  To improve our efficiency in
Virgo, we refined our selection criteria to a subsample of unresolved
(star-like) objects with colors similar to the Fornax objects.  The
targets were selected from the object catalog generated from United
Kingdom Schmidt Telescope plates by the Automated Plate Measuring
Machine (APM) in Cambridge (Irwin, Maddox \& McMahon 1994; see also
Maddox et al., 1990a,b).  Only objects classified by the APM as being
star-like (unresolved) or as merged (which includes star-like objects
merged with stars or with galaxies) were chosen, to exclude obvious 
galaxies.  Objects with physical sizes $\lesssim 100\:$pc (full-width 
at half-maximum) in the Virgo Cluster would be classified as 
unresolved. The apparent magnitude limits of the Virgo sample were set
to $b_j = 16.0$ to $20.2$~mag to match approximately the bright
absolute magnitude limit of the FCSS in the Fornax Cluster, while
extending about 1~magnitude fainter to probe more fully those objects
in the Virgo Cluster with luminosities similar to the nuclei of dE,N
galaxies.  The Virgo targets were further constrained to have color
indices $(b_j-r_F) \leq 1.6$~mag to limit their numbers, given the
colors already observed for the Fornax UCDs. Imposing this color 
restriction reduced the number of targets by about 30\% compared with 
a color-free sample. 

\begin{figure*}
\hspace*{-7mm}
\includegraphics[width=100mm,angle=0.]{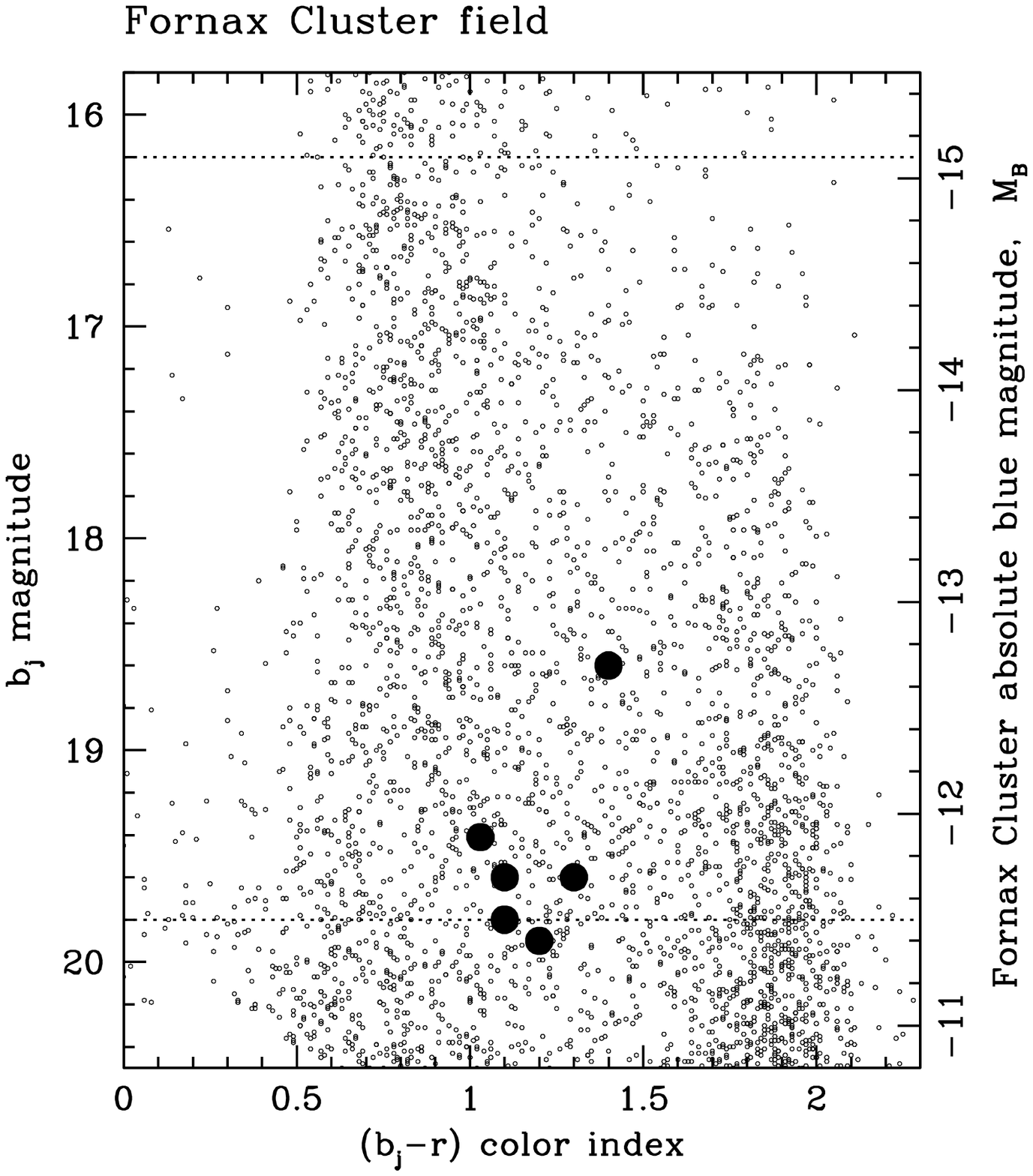}~\hspace*{-15mm}
\includegraphics[width=100mm,angle=0.]{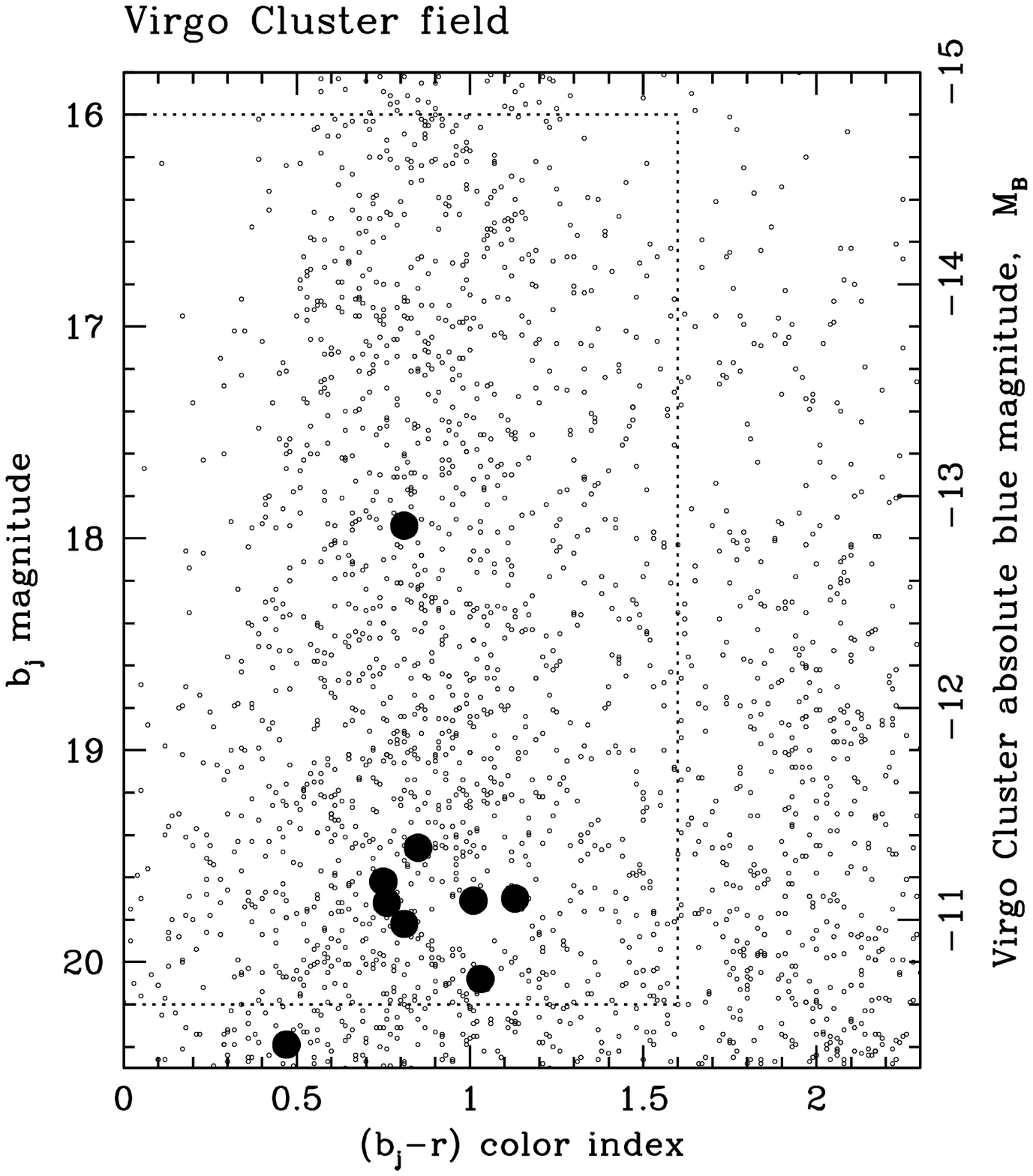}
\vspace*{-7mm}
\caption{ The color--magnitude diagrams for star-like objects in the
Fornax and Virgo Cluster fields taken from the APM Catalog.  Left:
the diagram for Field 1 of the FCSS.  The dotted lines show the 
nominal $b_j = 16.2$ and 19.8 magnitude limits on the FCSS star-like 
sample. The Fornax 
UCDs are shown as large solid circles.  Right: the color--magnitude
diagram for the 2dF field centered on M87 in the Virgo Cluster.  The
limits in color index and magnitude of the targets for the Virgo
compact galaxy survey are shown by dashed lines.  The Virgo UCDs
described in Section~\ref{sec_virgoresults} are shown as large solid
circles.  }
\label{fig_virgocmd}
\end{figure*}

A more complete sampling of targets was possible within a $0\fdg65$
radius of M87 ($72\,$\% of the objects meeting the magnitude, colour 
and type criteria) 
and a sparser sampling ($43\,$\% in the outer part of the 2dF
field (imposed by observing time constraints).  
This strategy produced a sample of 1501 targets in the M87 field; 
Figure~\ref{fig_virgocmd} presents the color--magnitude diagrams for
both the Fornax and the Virgo fields.  The limits in color and
magnitude used to define the two sets of targets are shown.  The Virgo
magnitude limits correspond to blue absolute magnitudes of $M_B =
-10.6$ to $-14.8$~mag at the distance of the cluster.  Although the
areas of the Fornax and Virgo fields are identical, the respective
numbers of stars differ because Fornax lies at a lower Galactic
latitude, $b \simeq -54^\circ$, than Virgo, $b \simeq +74^\circ$.

\begin{figure*}
\hspace*{15mm}
\includegraphics[width=130mm,angle=0.]{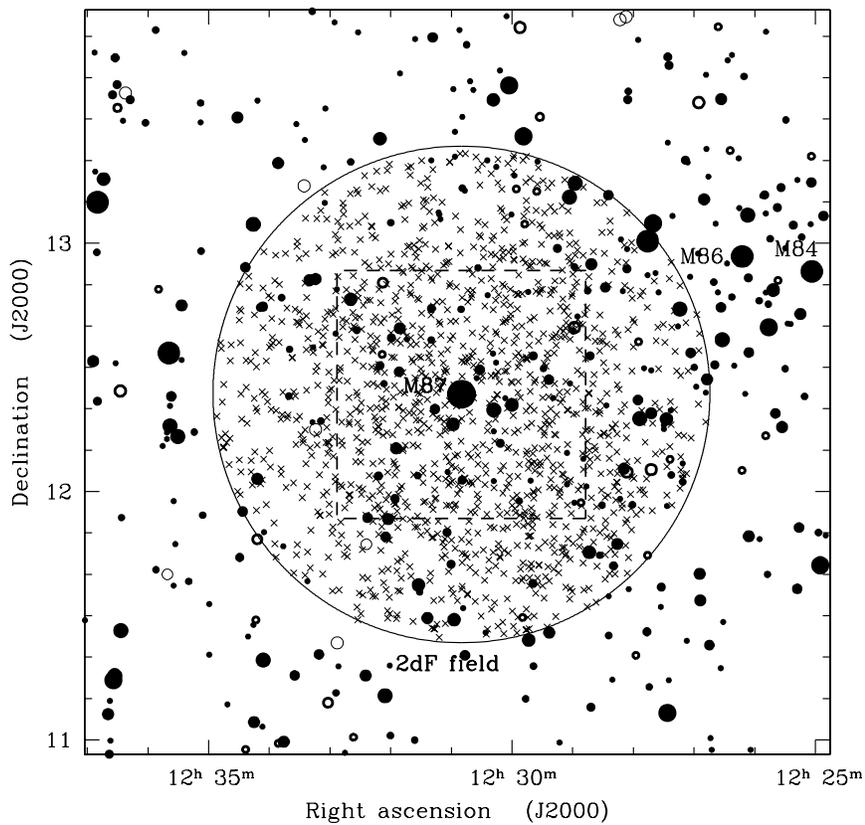}
\caption{
The 2dF spectrograph field of the Virgo Cluster ultra-compact dwarf 
survey. Galaxies from the Virgo Cluster Catalog of Binggeli, 
Sandage \& Tammann (1985) are plotted, with cluster members shown as 
solid circles, possible cluster members as thick open circles, 
and background galaxies as thin open circles. 
The sizes of the galaxy symbols are scaled by apparent magnitude. 
2dF targets are shown as small crosses. 
The dashed square delineates the region shown in Figure~\ref{fig_virgoucds}.
}
\label{fig_vcc}
\end{figure*}

\begin{figure*}
\vspace*{-25mm}
\includegraphics[width=180mm,angle=0.]{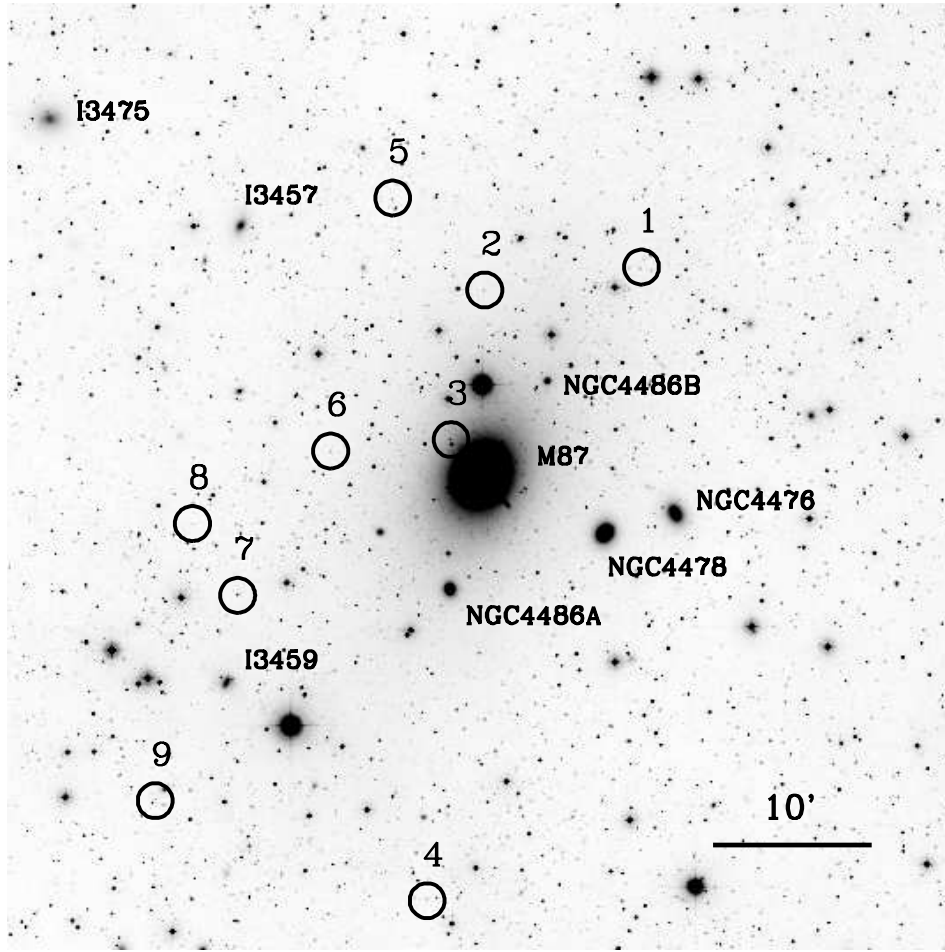}
\vspace*{-28mm}
\caption{
The inner 1.0~deg square region of the 2dF Virgo field, centered 
on M87, showing the positions of the 
nine ultra-compact dwarfs identified in the survey 
(labelled 1 to 9). North is at the top. 
The figure is based on a SuperCOSMOS scan of a UKST red Tech Pan 
film. 
}
\label{fig_virgoucds}
\end{figure*}

\subsection{Observations and 2dF data reduction}
\label{sec_virgoobs}

\begin{table}
\caption{Virgo Cluster 2dF observations}
Field center coordinates: \\[1mm]
\begin{tabular}{lll}
 & Right ascension & ~~$\,12^{\rm h}$  30$^{\rm m} $ 50$^{\rm s}$  (J2000) \\
 & Declination     & +12$^{\circ}$  23\farcs5   \\
\end{tabular}
\mbox{ } \\[3mm]
%
\begin{minipage}{60mm}
Exposures on 2dF target set ups:\\
\begin{tabular}{llll}
 & 2002 April 11 & 55 min total & $1.8''$ seeing\\
 & 2002 July 5   & 30 min       & $1.6''$  \\
 & 2002 July 6   & 90 min       & $1.4''$  \\
 & 2004 March 15 & 90 min       & $1.5''$  \\
 & 2004 March 15 & 80 min       & $1.5''$  \\
\end{tabular}
\end{minipage}
\label{tab_virgoobs}
\end{table}

AAT service observations were obtained on five separate nights 
for five separate 2dF set ups, totalling 6~hours of integration time. 
The observations are summarized in Table~\ref{tab_virgoobs}. 

The 2dF spectra were obtained with the 300B grating centered at
5800{\AA} and cover the wavelength range 3600 to 8000{\AA} at a
resolution of 9{\AA} (FWHM).  The signal-to-noise ratio is at least 10
per resolution element. The data were reduced with a combination of
IRAF and the standard Anglo-Australian Observatory 2DFDR utility
(Bailey, Glazebrook \& Bridges 2002).  As is normal for 2dF data, arc
lamp exposures and fiber flats were recorded, but no sky frames or 
dark-current exposures.  Each spectrum was wavelength calibrated but
not flux calibrated.  After reduction with 2DFDR, IRAF was used to
correct any spectra suffering from poor quality sky background
subtraction after the automated reduction by 2DFDR.  
IRAF was also used to remove cosmic ray event
detections.  Splitting integrations into a few different exposures
allows most cosmic rays to be eliminated from the coadded data.

\subsection{Data analysis}
\label{sec_2dfreduction}

Radial velocities were determined through cross-correlation with 
an array of template spectra representing an emission-line galaxy, 
a quasar, and nine different types of star (having spectral types 
from A3V to M5V to provide a match both to stars and early-type galaxies). 
This was done within the RVSAO IRAF package (Kurtz \& Mink 1998). 
The template giving the highest $R$ coefficient (Tonry \& Davis 1979) 
provides the adopted radial velocity for an observed spectrum, 
as well as a simple indication of the type of spectrum. 
Further details of the observing and reduction were 
given in Drinkwater et al.\ (2000b) and Deady et al.\ (2002). 

When a target has been observed more than once, which does sometimes 
occur because of fiber positioning practicalities, the result having 
the highest $R$ coefficient is adopted. Only results having $R \geq 3.0$ 
are considered reliable.

\subsection{Discovery of ultra-compact dwarfs in Virgo}
\label{sec_virgoresults}

\begin{deluxetable}{lcllcccrc}
\rotate
\tablecaption{New 2dF ultra-compact dwarfs identified in the Virgo and Fornax Clusters}
\vspace*{4mm}
\startdata
\multicolumn{1}{c}{Object} & IAU name & \multicolumn{1}{c}{R.A.}  &  \multicolumn{1}{c}{Dec.}  & $b_j$ & 
$(b_j-r)$ & $R$ & $cz$~~~~~ &  Best \\
 & &  ~~$^{\rm h}$~~$^{\rm m}$~~~$^{\rm s}$ & ~~~$^\circ$~~~$'$~~~$''$ & mag & mag & coeff.& (kms$^{-1}$)~ & template   \\[1mm]
Virgo UCD 1 & J123007.6+123631 & 12 30 07.61 & +12 36 31.1 &  19.6 & 0.8 & 5.2 & $ 1103 ~\pm \, 114$ & M1V \\
Virgo UCD 2 & J123048.2+123511 & 12 30 48.24 & +12 35 11.1 &  19.7 & 1.0 & 7.4 & $  824 ~\pm ~~ 50$ & K5V \\
Virgo UCD 3$^1$ & J123057.4+122544 & 12 30 57.40 & +12 25 44.8 &  19.5 & 0.9 & 9.9 & $  698 ~\pm ~~ 40$ & K5V \\
Virgo UCD 4 & J123104.5+115636 & 12 31 04.51 & +11 56 36.8 &  19.7 & 0.8 & 8.0 & $  824 ~\pm ~~ 46$ & G0V \\
Virgo UCD 5 & J123111.9+124101 & 12 31 11.90 & +12 41 01.2 &  19.7 & 1.1 & 7.3 & $ 1159 ~\pm ~~ 45$ & K5V \\
Virgo UCD 6 & J123128.4+122503 & 12 31 28.41 & +12 25 03.3 &  19.8 & 0.8 & 6.4 & $ 2041 ~\pm ~~ 63$ & M1V \\
Virgo UCD 7 & J123152.9+121559 & 12 31 52.93 & +12 15 59.5 &  17.9 & 0.8 & 6.1 & $  790 ~\pm ~~ 65$ & K5V \\
Virgo UCD 8 & J123204.3+122030 & 12 32 04.36 & +12 20 30.7 &  20.4 & 0.5 & 3.5 & $ 1607 ~\pm ~~ 68$ & K5V \\
Virgo UCD 9 & J123214.6+120305 & 12 32 14.61 & +12 03 05.4 &  20.1 & 1.0 & 4.9 & $ 1216 ~\pm ~~ 61$ & K5V \\[4mm]
Fornax UCD 6 & J033805.0$-$352409 & 03 38 05.08 & $-35$ 24 09.6 & 19.4 & 1.0 & 11.2 & $1212 ~\pm~~32$ & K5V  \\[2mm]
\enddata
\label{tab_fornaxucds}
\mbox{ }\\[2mm]
\begin{minipage}{200mm}
The properties of the new bright UCDs identified with 2dF are listed. 
The J2000 celestial coordinates are from the APM catalog. 
The APM $b_j$ magnitude is 
on the photographic blue (IIIa-J + GG395) photometric system. 
The APM photographic $(b_j-r)$ color index is accurate to $\pm 0.3\;$mag. 
$R$ is the Tonry \& Davis $R$ coefficient. 
$cz$ is the heliocentric radial velocity. The best template is 
the template spectrum used in the cross-correlation that gave the 
highest $R$ coefficient. \\[5mm]
$^1$ Virgo UCD 3 is object 547 in the study of the M87 globular 
cluster system by Strom et al. (1981).
\\[3mm]
\end{minipage}
\end{deluxetable}

\begin{figure}
\includegraphics[width=85mm,angle=0.]{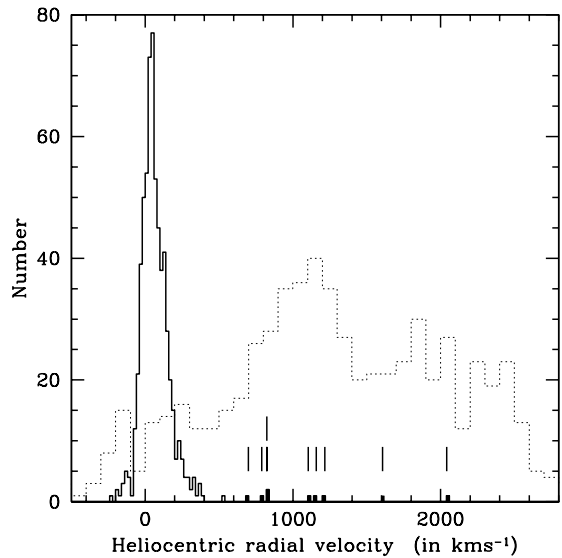}
\caption{
The histogram of velocities for the 2dF targets having $R \geq 3.0$ 
showing Galactic stars (solid line) and Virgo Cluster UCDs (solid areas, 
and also indicated by vertical lines). For comparison, the dotted 
line shows the velocity distribution for galaxies in the Binggeli, 
Sandage \& Tammann (1985) Virgo Cluster Catalog across the entire 
cluster using radial velocity data in the NASA Extragalactic Database. 
}
\label{fig_virgovelhist}
\end{figure}

A total of 1633 spectra were obtained of 1501 individual targets.
Of these, 1322 had reliable ($R \ge 3.0$)  velocity results. 
These results were obtained for 60\% of objects satisfying the 
object type, magnitude and color constraints over the whole 
2dF field, and for 70\% within 0\fdg65 of M87. 
Figure~\ref{fig_virgovelhist} presents the histogram of heliocentric
radial velocities for objects in the range $-500$ to 2200~kms$^{-1}$.
The distribution is dominated by Galactic stars between $-200$ and 
400~kms$^{-1}$.  Ten objects  have radial velocities
between 400 and 2100~kms$^{-1}$. One of these (at 12$^{\rm h}$ 31$^{\rm
m}$ 54{\fs}84, $+11^\circ$ 56' 59{\farcs}6, J2000) gave a velocity of
$527 \pm 121$~kms$^{-1}$ and its strongest cross-correlation result is
with the A-type stellar template.  This object is more likely to be 
a Galactic star, given the formal overlap of its velocity with the velocity 
distribution of Galactic stars and the relatively early spectral 
type of its best match template spectrum. 

About 10\% of already known Virgo members within our survey area have 
radial velocities $<400\;$kms$^{-1}$; some even have negative radial
velocities.  Thus a consequence of applying a low-velocity cut-off to
avoid Galactic star contamination in the compact object sample is that
the survey may miss UCDs with radial velocities below $400\;$kms$^{-1}$.
We estimate that there may be $1\pm 1$ additional UCDs among the 
objects observed that has been lost by applying this cutoff (assessed 
by applying this 
cut-off to the distribution of observed UCD velocities).  But UCDs with
$cz<~400\;$kms$^{-1}$ will be very difficult to distinguish from Galactic
stars, especially with poorly flux calibrated fiber spectra, so
obtaining a dynamically complete sample of Virgo UCDs will prove
difficult with currently available observing resources. 

\begin{figure}
\vspace*{-5mm}
\includegraphics[width=120mm,angle=0.]{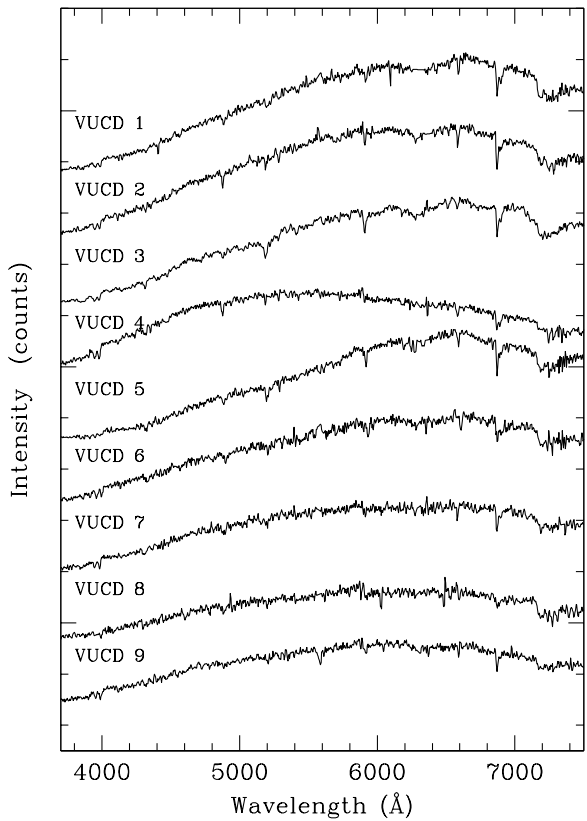}
\caption{
Spectra of the nine Virgo Cluster UCDs obtained with the 2dF 
spectrograph.
}
\label{fig_virgospectra}
\end{figure}

We identify as UCDs the remaining nine objects with velocities consistent 
with Virgo Cluster membership.  Like those in Fornax, all
nine Virgo UCDs have early-type galaxy spectra.  Most were fitted best
by stellar templates of K- or M-type solar-composition stars, 
consistent with old,
moderately metal-rich stellar populations.  The spectrum of Virgo
UCD~4, however, shows a bluer continuum
(Figure~\ref{fig_virgospectra}) and the best fitting template was that
of a G0-type star, which can be explained by stellar populations that
are younger or more metal-poor than the other UCDs.  
The objects are listed in Table~\ref{tab_fornaxucds}. 

A check of positions confirmed that none coincided with any known
cluster galaxies in the Virgo Cluster Catalog (Binggeli, Sandage \&
Tammann 1985), although UCD3 lies close to M87.  
None of the UCDs matches the objects described by Ha\c{s}egan et al. 
(2005), who studied specific fields within the Virgo Cluster. 
While there are Ha\c{s}egan et al. objects within our 2dF 
field, all their confirmed Virgo Cluster compact objects are located 
within 6' of M87, whereas the UCDs reported in this paper
lie between 3' and 29' of M87. 

Eight of the nine were classified in the APM Catalog
as being stars on both the blue and red sky survey plates. The
exception was UCD number 7 which was classified as a galaxy from the
red plate and a merged object on the blue.  This 
object appears on available imaging data to have a number of nearby
images which confuses its classification: this is discussed in more
detail in Sections \ref{sec_sbprofiles} and \ref{sec_lsbhalolimits}.

\subsection{Surface brightness profiles}
\label{sec_sbprofiles}

The UCDs are mostly star-like in the APM Catalog (one is a ``merged'' 
object). Better quality imaging data are therefore needed to attempt to 
resolve their surface brightness profiles.
Phillipps et al.\ (2001) found that one of five Fornax Cluster
ultra-compact dwarfs was resolved in Schmidt photographic images.
Using imaging from the Hubble Space Telescope, Drinkwater et al.\
(2003)
showed that this object (Fornax UCD~3) consists of a highly compact
core within a more extended envelope.  Resolving the Virgo UCDs is
important for comparing them to the Fornax sample and to further test
models of their origin.  Bekki et al.\ (2003), for instance, predict
that under some circumstances in the dE,N tidal threshing model,
residual halos can remain around UCDs.

We have used CCD data from the Isaac Newton Telescope Wide-Field
Survey (INT WFS) of the Virgo Cluster (Trentham \& Hodgkin 2002) to
measure the surface brightness profiles of the Virgo UCDs.
Blue (B band) data have been selected because of the greater depth and
generally flatter background than the corresponding ultraviolet or
infrared data.  Each UCD has been imaged in more than one B band
exposure, so we have selected just the best-seeing B band exposures to
measure the light profiles.
Comparing the UCD surface brightness profiles with stars from the same
images shows no strong evidence that any of the UCDs are resolved,
although UCDs 4 and 7 hint at being marginally extended in the profiles. 
Table~\ref{tab_virgoucdintsizes} summarises the object size data. 
UCD7 appears extended in the wings of its profile only and therefore 
does not appear extended in Table~\ref{tab_virgoucdintsizes}.

\begin{table}
\caption{Images sizes for the Virgo UCDs in INT CCD data}
\begin{tabular}{ccc}
 & & \\
Virgo UCD & Frame seeing ($''$) &   UCD size ($''$)  \\
          &  ($\pm0\farcs1$)     &   ($\pm0\farcs2$)   \\[2mm]
    1     &     1.8      &   1.8     \\
    2     &     1.4      &   1.6     \\
    3     &     1.1      &   1.3     \\
    4     &     1.9      &   2.4     \\
    5     &     1.1      &   1.2     \\
    6     &     1.0      &   1.2     \\
    7     &     1.1      &   1.3     \\
    8     &     1.5      &   1.1     \\
    9     &     1.6      &   1.8     \\[3mm]
\end{tabular}
\mbox{ } \\
\label{tab_virgoucdintsizes}
The object sizes of the UCDs in CCD data from the Isaac Newton Telescope 
are compared with the mean sizes of stars. 
\mbox{ } \\[3mm]
\end{table}

\begin{figure*}
\vspace*{0mm}
\hspace{0mm}
\includegraphics[width=160mm,angle=0.]{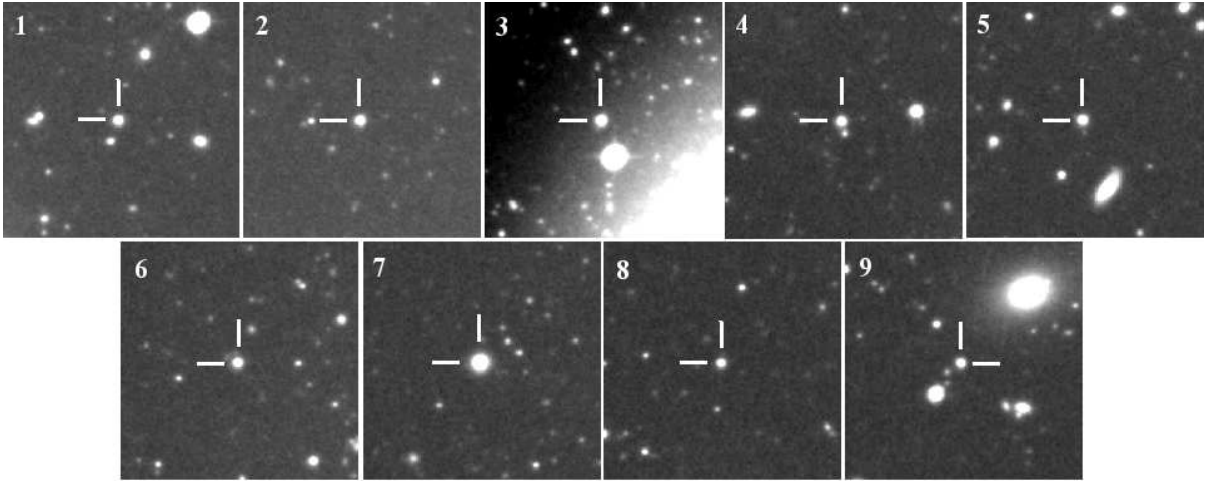}
\caption{
Images of the nine Virgo Cluster UCDs from the stack of 63 UKST red band 
films. Each frame is 2\farcm0 wide and north is at the top. The UCD 
is centered in each frame. 
}
\label{fig_virgoucdpics}
\end{figure*}

Figure~\ref{fig_virgoucdpics} shows images of the Virgo UCDs created
from a very deep stack of photographic films of the M87 region
recorded with the UKST.  The data were produced by the SuperCOSMOS Sky
Unit at the Royal Observatory Edinburgh by coadding 
digitized scans of 63 red filter exposures on
Tech Pan emulsion (see Phillipps et al.\ 1998 for a discussion of 
this type of data).  The point-spread function in the coadded data is
$2.5\arcsec$, poorer than the available INT CCD data; however, the
photographic data have a much greater depth and 
uniform quality.  These data are discussed in further detail in
Sections~\ref{sec_lsbhalolimits} and~\ref{sec_lsbdens}.

Virgo UCD 6 has three compact objects $6\arcsec$ just to the north, 
northeast and east in the INT B band and infrared (i band) images: 
see Figure~\ref{fig_intwfs_vucda6}, which presents a B-band image 
of UCD~6 recorded in a 25-minute integration with the Wide-Field 
Camera on the 2.5-metre Isaac Newton Telescope in La Palma. 
These three objects lie 6 to $7\arcsec$ from the UCD, which
corresponds to a projected distance of about 500~pc in the Virgo
Cluster.  Two of them are compact, while the third is resolved and 
is elongated.  They appear as two objects in the lower resolution 
UKST photographic data.  It is not clear from the imaging data 
whether these are three background/foreground objects or whether 
they are possible debris from the tidal stripping of a progenitor 
galaxy (see also Sections~\ref{sec_lsbhalolimits} and~\ref{sec_discuss}). 
The 2dF spectrum, given the fiber-fed character of 2dF, included 
light from only the UCD and not the companion objects. The compact 
images appear bluer by 0.3~mag than the UCD in (B-i), while the 
elongated image is redder by 0.4~mag. 
There is no obvious evidence of discrete structures around any of the 
other UCDs. Section~\ref{sec_lsbhalolimits} below discusses limits on 
halos around the UCDs.

\subsection{The luminosities of the UCDs}

\begin{figure}
\vspace*{-8mm}
\hspace{-5mm}
\includegraphics[width=85mm,angle=0.]{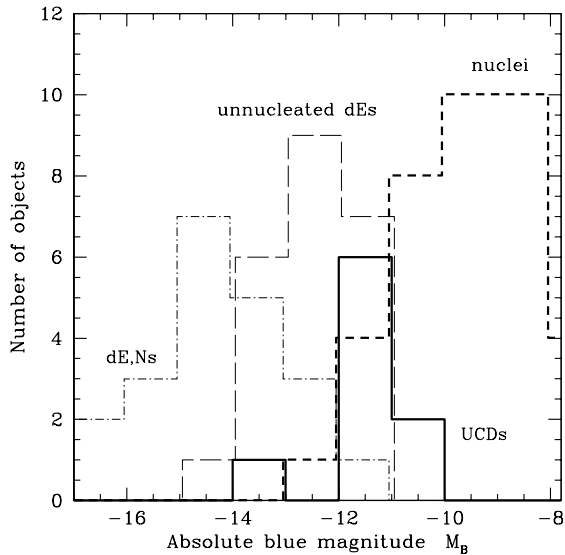}
\caption{ The luminosity function of the Virgo UCDs compared with
other objects in the Virgo Cluster.  The distribution of absolute
magnitudes of the UCDs is shown as a bold solid line.  Total absolute
magnitudes of nucleated dwarf elliptical (dE,N) and non-nucleated
dwarf elliptical galaxies were taken from data in the Virgo Cluster
Catalog (Binggeli, Sandage and Tammann 1985). These are presented
for those galaxies lying within $0\fdg6$ of M87 only, to allow a direct
comparison with the UCDs.  The absolute magnitudes of the nuclei of
Virgo dE,N galaxies are taken from Lotz, Miller \& Ferguson (2004) and from
Durrell (1997).  
}
\label{fig_lfvirgoucds}
\end{figure}

Figure~\ref{fig_lfvirgoucds} presents a histogram of the B band
absolute magnitudes for the UCDs, as computed from the apparent
magnitudes using an apparent (reddening uncorrected) 
Virgo Cluster distance modulus of 31.0~mag.  Note that absolute 
magnitudes are expressed here for the B band, converted from the 
$b_j$ photographic apparent magnitudes using a color typical 
of evolved stellar populations. 
The absolute magnitudes range from $M_B = -12.9$ to
$-10.7$, while the survey limits were $-14.8 \leq M_B \leq -10.6$.
These are comparable to the luminosities of the six bright Fornax UCDs 
($M_B = -13.8$ to $-11.6$, with survey limits of $-15.1$ to $-11.5$). 
In both clusters, the UCDs so far identified have luminosities towards 
the fainter end of the range explored with 2dF.  This suggests that 
the peak in the Virgo UCD luminosity function could lie at $M_B > -12$, 
and possibly fainter than the faint limits of the survey, predicting 
larger numbers of these objects at fainter levels (as has been found 
to be the case in Fornax), subject to confusion with any intracluster 
globular clusters. .

Figure~\ref{fig_lfvirgoucds} also shows the luminosity functions of
other objects in the Virgo Cluster.  Comparisons are possible with the
nuclei of dE,N galaxies, which overlap with the UCD luminosities at
the luminous end of the nucleus distribution.  This result is similar
to that of Drinkwater et al.\ (2000a) for Fornax UCDs.  A direct
comparison of the numbers of UCDs and Virgo dE,N nuclei cannot,
however, be made because of the heterogeneous samples of dE,N nuclei 
that have reliable nuclear magnitudes in the literature.
Figure~\ref{fig_lfvirgoucds} shows the total luminosities of 
dE,N and non-nucleated dE galaxies within the same projected radial 
distance ($0\fdg6$) of M87 as the UCDs.  The numbers of dE,Ns, dEs and 
UCDs can be compared directly.

\subsection{The distribution of the UCDs within the Virgo Cluster}
\label{sec_virgoradialdistn}

Figure~\ref{fig_raddistn} presents the cumulative distribution
of the Virgo UCDs as a function of radial angular distance from M87.
All of the Virgo UCDs lie within $0\fdg6$ of M87 (projected distance
of 170~kpc), though this might be due in part to the sparser sampling of
targets outside this radius.  Half of the UCDs lie within $0\fdg3$
(80~kpc) of M87.

Figure~\ref{fig_raddistn} also compares the angular distribution 
of the UCDs with those of non-nucleated and nucleated 
dwarf ellipticals (taken from the Virgo Cluster Catalog of Binggeli et al.\ 
1985), normalized to match the number of UCDs within $0\fdg5$ of M87 
(the maximum radial distance found for the UCDs). There is a moderate 
match between the UCD distributions and the two types of dwarf ellipticals. 
However, dwarf ellipticals are found across the cluster and this 
result depends critically on the radius used for the normalization. 
A Kolmogorov-Smirnov test shows that the distribution of non-nucleated 
dwarf ellipticals normalized within $0\fdg5$ of M87 matches that of 
the UCDs with a 32\% probability. The sparser sampling of 2dF targets 
beyond $0\fdg6$ of M87 does not affect this result. 

A direct comparison of the distribution of UCDs with the M87 globular 
cluster system is more difficult given the restricted range in radii 
covered by past surveys of the M87 globular cluster system. For example, 
the study of McLaughlin et al. (1994) covers radii out to $9\arcmin$: 
in that study, 50\% of globular clusters are found within $3\arcmin$ of 
the centre of M87. 
In contrast 50\% of the Virgo UCDs of Table~\ref{tab_fornaxucds} lie 
within $17\arcmin$ of M87, while only one of the nine lies within $9\arcmin$.

\subsection{Virgo UCD dynamics}
\label{sec_radvels}

The mean heliocentric radial velocity of the nine UCDs is $1051 \pm
139 \; \mbox{km} \mbox{s}^{-1}$ (weighted according to the errors
given by the RVSAO software), with a dispersion of $416 \pm 110 \;
\mbox{km} \mbox{s}^{-1}$.  Interpretation of these results is
slightly complicated by the lower velocity cut-off at $400 \; \mbox{km} \,
\mbox{s}^{-1}$ that was imposed on the compact-object sample to reject
Galactic stars. 

The sample of Virgo UCDs is too small for definitive conclusions 
to be made about whether their systemic velocity is consistent 
with that of M87  (having a heliocentric radial velocity of 
$1307 \pm 7 \; \mbox{km} \,\mbox{s}^{-1}$ 
in the NASA Extragalactic Database, NED)\footnote{
The NASA/IPAC Extragalactic Database is operated by the Jet
Propulsion Laboratory, California Institute of Technology, under
contract with the National Aeronautics and Space Administration.},
with that of dE,N galaxies in the vicinity of M87, or with that of 
the M87 globular cluster system. 
Differences in the mean velocity and the velocity dispersion 
might provide tests of the origin of the UCDs.
The mean velocity and dispersion of the UCDs are statistically 
different to the entire sample of cluster galaxies close to M87 
(there are 37 galaxies with NED velocities between 400 and 
$3500 \:\mbox{km}\,\mbox{s}^{-1}$ lying within
$1\fdg0$ of M87; these have a mean
velocity of $1402 \pm 105 \; \mbox{km} \mbox{s}^{-1}$ and a velocity
dispersion of $639 \pm 75 \; \mbox{km} \mbox{s}^{-1}$).  
The larger velocity dispersion of the general galaxy population than 
that of the UCDs parallels the results for Fornax UCDs by Mieske, 
Hilker \& Infante (2004);  but this conclusion depends on imposing 
a lower velocity limit of $400 \:\mbox{km}\,\mbox{s}^{-1}$ on 
cluster galaxies so that the samples have similar velocity selection 
criteria.

\subsection{Limits on low surface brightness halos around the UCDs}
\label{sec_lsbhalolimits}

Phillipps et al.\ (2001) showed that one of their Fornax UCDs was
resolved in ground-based imaging.  Using numerical simulations, Bekki
et al.\ (2003) showed that dE,N nuclei might be only partially
stripped in their threshing model, leaving a halo around a UCD.  The
existence of faint halos around any of the Virgo UCDs might
provide important information about formation mechanisms.

The deep stack of UKST photographic films produced by the 
SuperCOSMOS Unit, described in Section~\ref{sec_sbprofiles}, 
was used to place limits on emission within $8\arcsec$ of the 
UCDs, above that expected from a point-spread function. 
The pixel-to-pixel variation in the sky background is at the 
26.3 R~mag arcsec$^{-2}$
level and R = 23.5~mag point sources are detected with a
signal-to-noise ratio of 5.  A circular annulus having an inner radius
of $5\arcsec$ and an outer radius of $8\arcsec$ (corresponding to 
a transverse linear distances of 390 and $620\:$pc in the Virgo 
Cluster) was centered on each
UCD and the signal in the annulus recorded. The inner radius is set by
the point-spread function: 6\% of the light of a point source
spills into this annulus (measured from images of stars).  
The full-width at half-maximum of the
point-spread function is $2.5\arcsec$;  this figure is uniform across 
the region where the UCDs are located. Some discrete objects in the
vicinity of the UCDs were masked before making the measurements.  This
did not include the objects very close to UCD 6, and the procedure may
have been incomplete for UCD 7 which had several objects within
$20\arcsec$.

\begin{table}
\caption{Limits on halos around the Virgo UCDs}
\begin{tabular}{ccc}
  Virgo UCD & $2.5\sigma$ limit     & Mean surface brightness     \\
            & above sky level       & if above $2.5\sigma$ limit  \\
            & (R mag arcsec$^{-2}$) & (R mag arcsec$^{-2}$) \\[2mm]
      1     &         27.5          &    --      \\
      2     &         27.5          &    --      \\
      3     &         26.7          &    --      \\
      4     &         27.3          &    --      \\
      5     &         27.3          &    --      \\
      6     &         27.4          &    $26.4 \pm 0.2$     \\
      7     &         26.3          &    $24.7 \pm 0.1$     \\
      8     &         27.8          &    --      \\
      9     &         27.5          &    --      \\[3mm]
\end{tabular}
\mbox{ }\\
\label{tab_virgoucdhaloes}
The mean surface brightnesses and limits are presented for 
circular annuli with inner and outer radii of 5 and 8~arcsec 
centered on the UCDs. The results represent the signals in the 
annuli after corrections for light spilling from the point-spread 
function of the UCDs. 
\mbox{ } \\[3mm]
\end{table}

Table~\ref{tab_virgoucdhaloes} presents $2.5\sigma$ limits on the detection of 
emission in the annuli above that from the point-spread function. 
These are expressed as the mean magnitude surface brightness 
in the annuli corresponding to a $2.5\sigma$ deviation 
above the sky background 
level after applying a correction for light spilling from the UCD. 
A significant contribution to the error is the uncertainty 
in the point-spread function ($5.6 \pm 1.1 \%$ of the UCD light is 
expected to spill into the annulus, based on tests on images of stars). 

There is a formal detection of excess light around two UCDs. However,
one of these (number~6) has the three discrete contaminating sources
within $6\arcsec$ from the UCD center (discussed in
Sections~\ref{sec_sbprofiles} and~\ref{sec_discuss}).  The other
(number~7), as discussed above, also has several additional sources
within $20\arcsec$ and some within $10\arcsec$.  Although these
objects were masked, some residual signal may still contaminate the
annulus.  It is therefore not possible to claim with confidence any
evidence of extended halos around any of the Virgo UCDs.

\subsection{Might the UCDs be the nuclei of very low surface 
   brightness dE,Ns?}
\label{sec_lsbdens}

\begin{figure}
\vspace*{-8mm}
\hspace{-5mm}
\includegraphics[width=85mm,angle=0.]{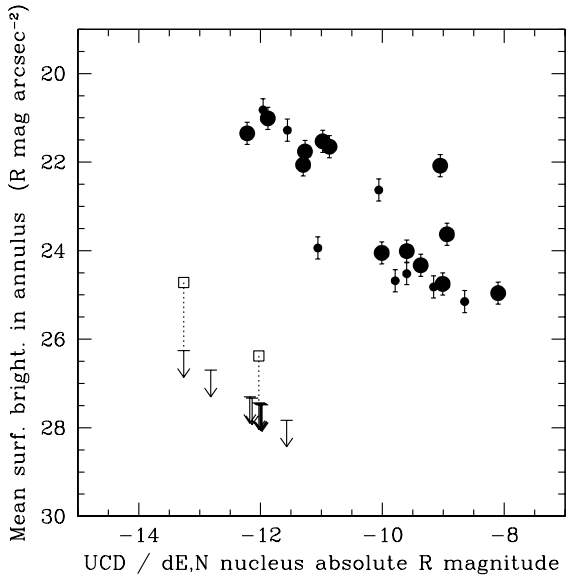}
\caption{ A comparison of the 2.5$\sigma$ surface brightness limits on
extended halos around the Virgo UCDs with the detections of dE,N
galaxies when the same analysis is performed on dE,N nuclei.  The
2.5$\sigma$ limits on the signal in the circular annuli of
Section~\ref{sec_lsbhalolimits} are plotted against the UCD absolute
magnitudes and are shown as arrows. The two UCDs (numbers 6 and 7)
that gave formal detection are shown by squares, alongside their
respective 2.5$\sigma$ limits: the reliability of these two detections
are discussed in Section~\ref{sec_lsbhalolimits}.  Equivalent data for
Virgo (large circles) and Fornax (small circles) dE,N galaxies are
shown for comparison, in the form of the mean surface brightness in 5
to 8~arcsec radius annuli centered on the nucleus plotted against
nucleus absolute magnitude. Section~\ref{sec_lsbdens} discusses the
comparison of UCDs with dE,Ns in more detail.  }
\label{fig_virgoucds_dENnucs}
\end{figure}

It is important to rule out the possibility that the UCDs are merely
the high surface brightness, compact nuclei of dE,N galaxies with
lower than average surface brightness halos.  Phillipps et al.\ (2001) and
Drinkwater et al.\ (2003) compared limits on the surface brightnesses
of extended light around Fornax UCDs in relation to surface
brightness data for dE,N galaxies.  Their analyses showed that their
five Fornax UCDs were different in character to ordinary dE,N
galaxies.

This analysis is repeated here for the nine Virgo UCDs.  The 5 to
$8\arcsec$ annuli used in Section~\ref{sec_lsbhalolimits} were applied
to a sample of Virgo dE,N galaxies, using the same stack of UKST films
as for the UCD analysis. This provided mean surface brightnesses
within 5 to $8\arcsec$ of the nuclei of the dE,N galaxies: all are
easily detected.  Figure~\ref{fig_virgoucds_dENnucs} shows the mean
surface brightness in the annuli plotted against the magnitude of the
nucleus of the dE,N.  Errors are estimated to be $\pm 0.2\;$mag. 
We used the best nuclear magnitude estimates
from the literature, derived from high-resolution imaging (see the
discussion of Binggeli \& Cameron 1993 about the difficulties of
measuring nuclear magnitudes from photographic data).  The nuclear
magnitudes from the HST WFPC2 study by Lotz, Miller \& Ferguson (2004) 
were converted to R-band absolute magnitudes (for our 
adopted Virgo distance modulus of 31.0~mag, and using a transformation 
$(V-R)_{_C} = 0.52 \, (V-I)_{_C}$ based on data given by Bessell (1979) for 
solar-composition main sequence stars).  One dE,N nucleus was
rejected from this analysis: VCC1714 has a nuclear color from 
Lotz, Miller \& Ferguson (2004) 
that is 5.3 standard deviations from the mean of the other 29
nuclei in their  study, and it is displaced off-center to an
unusual degree in the galaxy image.  We reject VCC1714 on the basis
that it is an unusual object or that its ``nucleus'' may be a Galactic 
star (but see, for example, De Rijcke \& Debattista 2004 for a 
discussion of off-center nuclei).   This sample of Lotz, Miller \& 
Ferguson galaxies was supplemented by data for one galaxy (VCC1783) 
from Lotz et al.\  (2001), and from the ground-based CCD study
of Durrell (1997)
for 3 galaxies, giving a total of 14 Virgo dE,Ns within the region 
of the SuperCOSMOS scans. 

These Virgo dE,N galaxy results were supplemented by an analysis 
of a SuperCOSMOS 
scan of a single UKST film on Tech Pan emulsion of the center 
of the Fornax Cluster.  This provided data for a further 8 dE,Ns 
with nuclear magnitudes from Lotz et al.\ (2001). The Fornax 
results are shown as small circles in Figure~\ref{fig_virgoucds_dENnucs}, 
while the Virgo dE,Ns are shown as large circles. 
The Virgo galaxies show the same surface brightness -- nucleus 
magnitude correlation as the Fornax galaxies. 

Figure~\ref{fig_virgoucds_dENnucs} also shows the 2.5$\sigma$ limits 
for the non-detections of seven Virgo UCDs, with the formal detections 
for the other two UCDs. These are plotted against the R band absolute 
magnitudes of the UCDs, allowing a direct comparison to be made 
with the dE,N galaxies, in a similar manner to the Fornax analysis 
of Drinkwater et al.\  (2003). The correlation for the UCDs between surface 
brightness limit and magnitude is caused by the error in surface 
brightness introduced by uncertainties in the point-spread function. 
(In contrast, the contribution of nucleus light into the annulus 
around dE,Ns, at $0-8\:$\%, is negligible compared to the light of 
the main body of the galaxy.)   The limits for the UCDs are  
displaced from the locus of the dE,N galaxies by 5--6 magnitudes 
in surface brightness. On this basis we dismiss the hypothesis that 
UCDs are the nuclei of dE,N galaxies of lower than average surface 
brightness.

\begin{figure*}
\hspace*{-5mm}
\includegraphics[width=85mm,angle=0.]{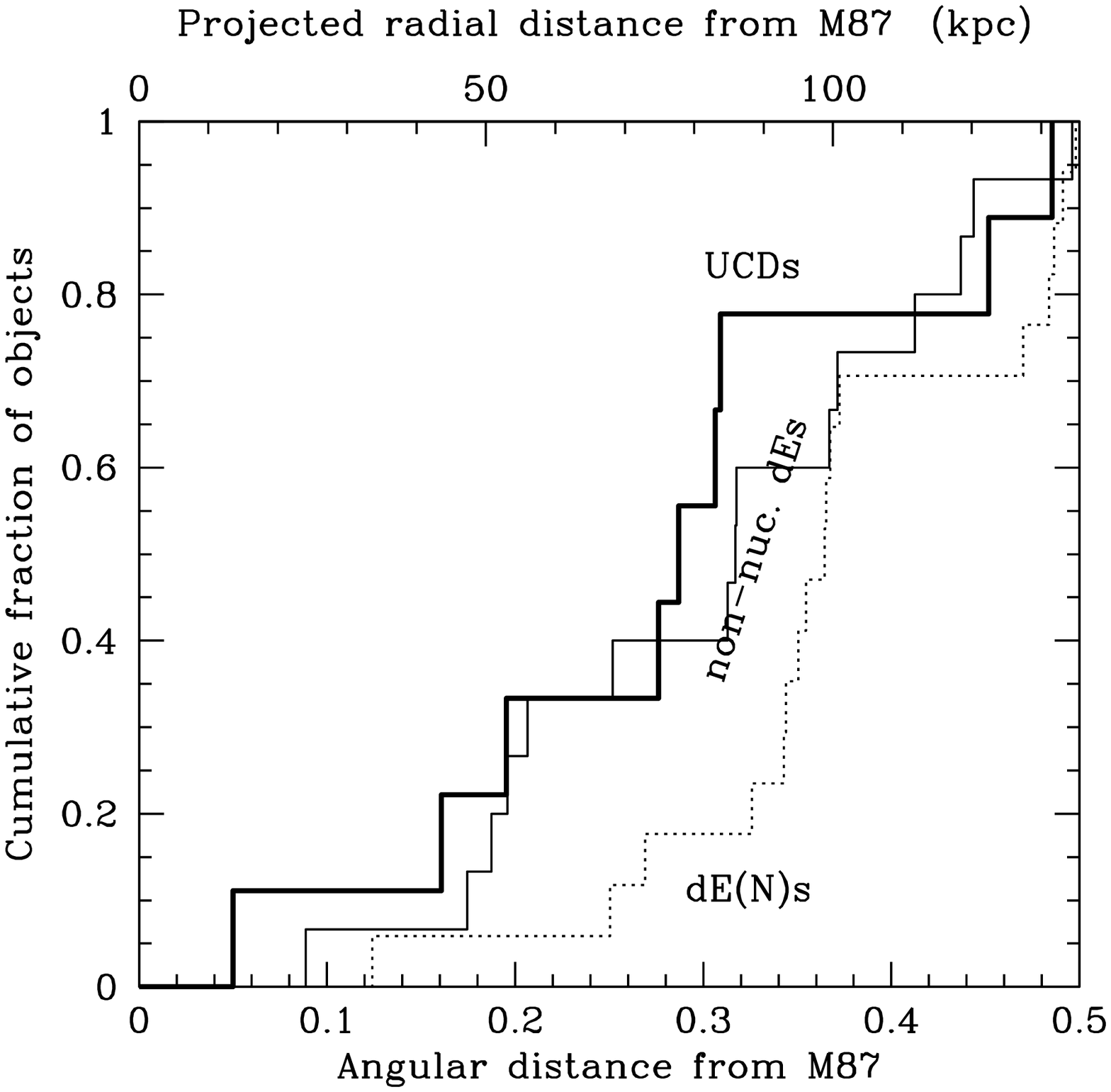}
\includegraphics[width=85mm,angle=0.]{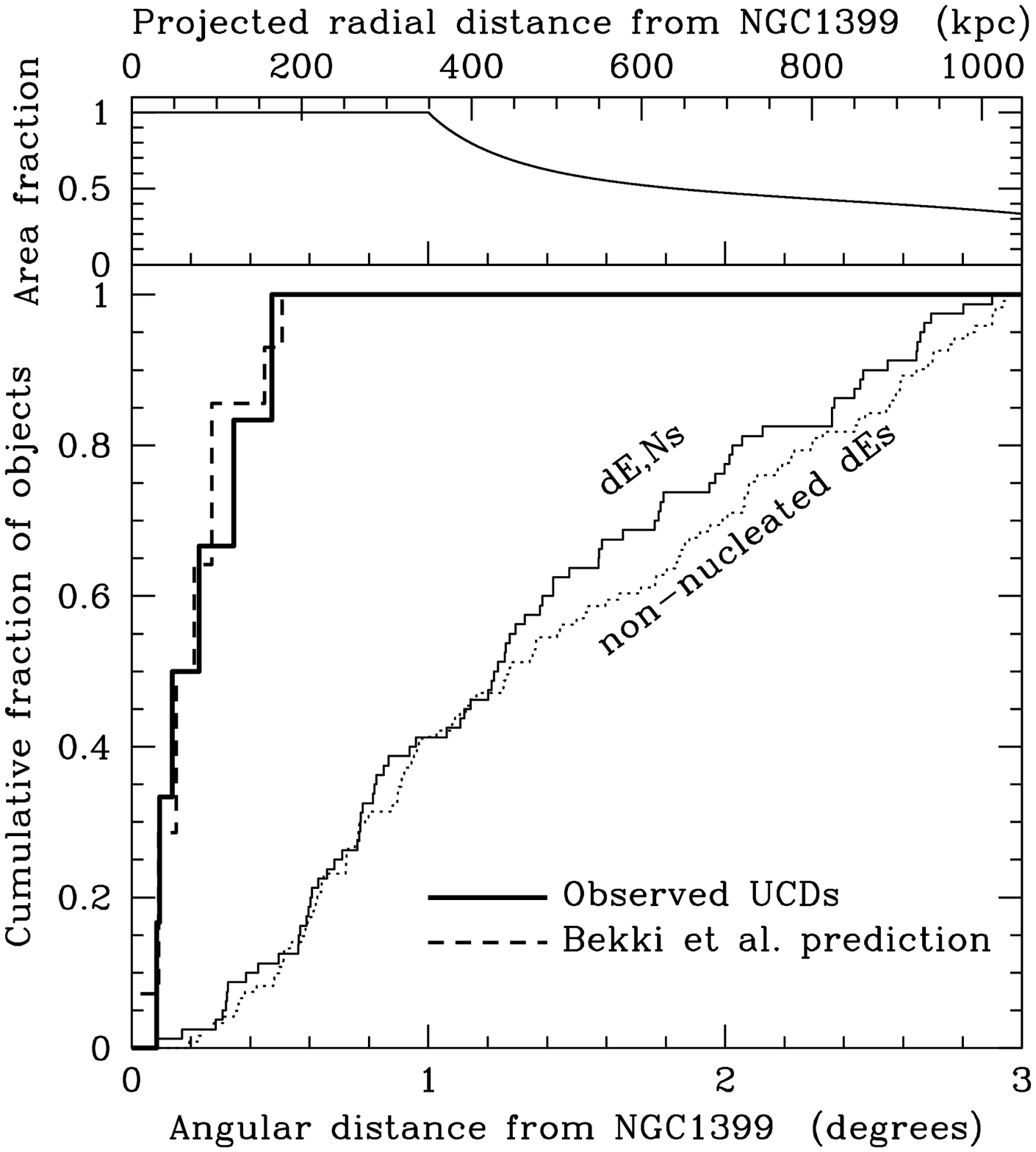}
\caption{
Left: The radial distribution of the nine  Virgo ultra-compact 
dwarfs. The number of UCDs found within a particular angular 
distance of M87 is plotted against the angular distance. 
The distributions of nucleated dwarf ellipticals 
(dE,Ns) and non-nucleated dwarf ellipticals are presented, taken 
from the Virgo Cluster Catalog of Binggeli, Sandage \& Tammann (1985) 
and normalized so that the numbers of galaxies within $0\fdg5$ of M87 
match the numbers of UCDs. 
Right: 
The radial distribution of the six Fornax ultra-compact 
dwarfs found in the FCSS, covering the full 3~degree radius of the survey. 
The cumulative number of UCDs found within a particular angular 
distance of NGC1399 is plotted against the angular distance, 
compared with the distribution of UCDs predicted by Bekki et al.\ (2003) 
on the basis of the tidal threshing model. 
The distributions of nucleated dwarf elliptical (dE,N) and 
non-nucleated dwarf elliptical galaxies are shown, taken from the 
Fornax Cluster Catalog of Ferguson (1989). 
The top panel shows the cumulative fraction of the area of sky surveyed 
by the FCSS as a function of distance from NGC1399. 
}
\label{fig_raddistn}
\end{figure*}

\section{Ultra-compact Dwarf Galaxies in the Fornax Cluster}
\label{sec_fornax}

\subsection{The Fornax Cluster Spectroscopic Survey}
\label{sec_fcss}

The Fornax Cluster Spectroscopic Survey (FCSS) has aimed to obtain
spectra with the 2dF spectrograph of all objects having blue
photographic magnitudes between $b_j = 16.5$ and 19.7 in the direction
of the Fornax Cluster (see Drinkwater et al.\ 2000b for a description
of the survey and Phillipps 1997 for the original strategy).  
Deady et al.\ (2002) have already described the 
results for dwarf galaxies in the cluster core from the first of the
2dF fields (for B-band absolute magnitudes mostly in the range $-11.6$
to $-14.8$).  The 2dF fields used in the survey are shown in
Figure~\ref{fig_fcc}.

The five UCDs reported by Drinkwater et al.\ (2000a) and Phillipps 
et al.\ (2001) were discovered in the first FCSS field at a time 
when 86\% of the star-like targets in that field had been observed
between magnitude limits at $b_j = 16.2$ and 19.8. The magnitude 
limits for the star-like targets was slightly broader than the nominal 
$b_j = 16.5$ to 19.7 limits of the overall survey. 
Since then the completeness of Field 1 has been improved by further 
observations and Fields 2 and 3 have been observed. 
Details of the fields are presented in Table~\ref{tab_fornaxobs}. 
Velocities have now been measured for 92\% of the Field 1 star-like 
objects to $b_j = 19.8$ and some targets have been observed to 
$b_j = 20.0$. 
The non-detection of UCDs in Fields 2 and 3 is discussed in 
Section~\ref{sec_fornaxucdsflds23}.

\subsection{Fornax UCD6: A bright ultra-compact dwarf in FCSS Field 1}
\label{sec_fornaxucds}

\begin{table}
\caption{New Fornax Cluster 2dF observations}
\vspace*{4mm}
Fornax Cluster Spectroscopic Survey fields : \\[2mm]
\begin{tabular}{lccc}
~ & Number  & \multicolumn{2}{c}{Field center coordinates (J2000)}  \\
  &         & Right ascension & Declination  \\[1mm]
& Field 1 & 03$^{\rm h}$ 38$^{\rm m}$ 29.0$^{\rm s}$ & $-35^{\circ}$ 27$'$  01$''$  \\
& Field 2 & 03$\;\;$     28$\;\;\,$   40.0$\;\,$     & $-35\;\:$     27$\;$ 01$\;\;$  \\
& Field 3 & 03$\;\;$     33$\;\;\,$   38.0$\;\,$     & $-33\;\:$     41$\;$ 59$\;\;$ \\[3mm]
\end{tabular}
\mbox{ }\\
\hspace*{15mm} Each field is 2\fdg0 in diameter.
\label{tab_fornaxobs}
\mbox{ } \\[3mm]
\end{table}

\begin{figure}
\vspace*{-15mm}\hspace*{-1mm}
\includegraphics[width=80mm,angle=0.]{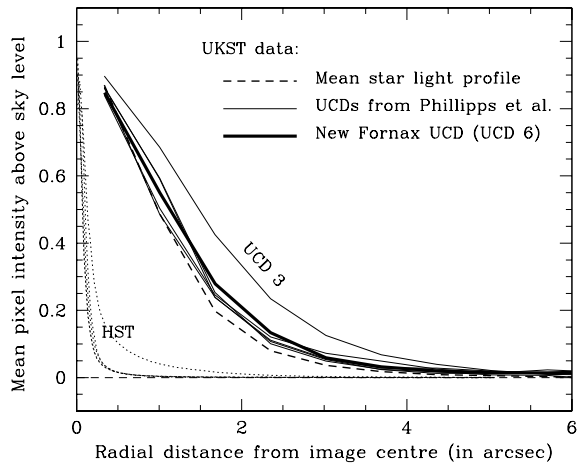}
\caption{
Radial light intensity profiles for Fornax Cluster UCD6 
derived from a SuperCOSMOS scan of a UKST red Tech Pan film. 
The profiles of the five UCDs of Drinkwater et al.\ (2000a) and a 
mean star point-spread function from the same UKST data 
are shown for comparison, taken from Phillipps et al.\ (2001). 
Resolved profiles from the Hubble Space Telescope
STIS data of Drinkwater et al.\ 
(2003) are shown for the first 5 UCDs 
to illustrate the effects of seeing on the UKST images. 
}
\label{fig_fornaxucdsprof}
\vspace*{15mm}
\end{figure}

\begin{figure*}
\hspace*{25mm}
\includegraphics[width=110mm,angle=0.]{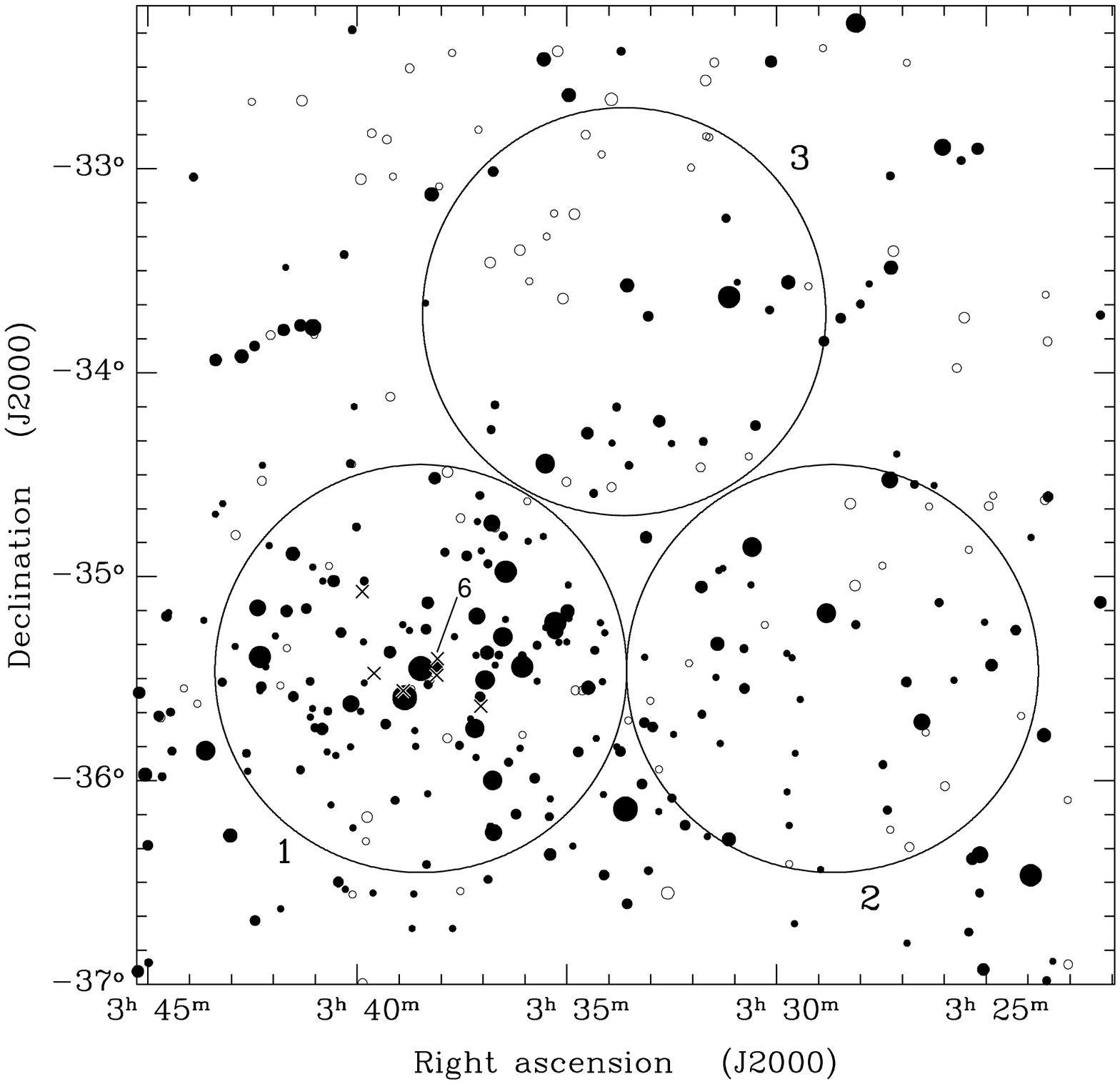}
\vspace*{0mm}
\caption{
Fields 1, 2 and 3 of the Fornax Cluster Spectroscopic Survey showing the 
galaxies from the 
Fornax Cluster Catalog of Ferguson (1989). 
The 2dF spectrograph fields of the Fornax Cluster Spectroscopic 
Survey (see also Drinkwater et al.\ 2000b) are drawn as circles 
and are numbered. The definite or likely cluster members 
of the Fornax Cluster Catalog are plotted as solid points, 
and possible cluster members as open circles. 
The sizes of the symbols are scaled by apparent magnitude. 
The six bright UCDs detected in the FCSS are shown as crosses, 
and UCD 6 is labelled. 
}
\label{fig_fcc}
\vspace*{5mm}
\end{figure*}

Only one UCD has been found in the Fornax Cluster from the FCSS observations 
in addition to the five reported by Drinkwater et al.\ (2000a) and 
by Phillipps et al.\ (2001). Its position and color have previously 
been described by Karick, Drinkwater \& Gregg (2003), who found 
V=18.9, (B-V)=0.8, (V$-$I)$_C$=1.2, and an independent velocity 
measurement and color have been reported by Mieske, Hilker \& Infante (2004).  
The properties of UCD6 are summarized in Table~\ref{tab_fornaxucds}. 
The object has a blue 
photographic apparent magnitude $b_j = 19.3$, equivalent to a blue
absolute magnitude $M_B = -12.0$, comparable to those of the first
five UCDs.  The best-fitting cross-correlation template spectrum is
that of a K5 star, again similar to the first five UCDs.  
UCD6 lies only 6\arcmin from the giant
elliptical galaxy NGC1399, equivalent to a projected distance of
35~kpc, whereas the entire distribution of bright UCDs extends as far as
28\arcmin (170~kpc).

Figure~\ref{fig_fornaxucdsprof} shows the radial light intensity 
profile of the new Fornax Cluster UCD determined from a SuperCOSMOS 
plate measuring machine scan of a red photographic film recorded 
with the United Kingdom Schmidt Telescope (UKST) on Technical Pan 
emulsion (Parker \& Malin 1999, Schwartzenberg, Phillipps \& 
Parker 1995). 
It is compared with the mean profile of star images 
from the same scan, and with the profiles for the 
five UCDs presented by Phillipps et al.\ (2001). 
The UCD is not obviously resolved in the photographic data 
(with a 2.3~arcsec seeing disc), unlike the third of the 
Drinkwater et al.\ (2000a) objects. 

Karick, Drinkwater \& Gregg (2003) published photometry for an 
additional object found during a preliminary analysis of 2dF data 
(which they labelled UCD7). A detailed analysis has failed 
to confirm this as a 
Fornax Cluster object. It appears to be a Galactic star which gave 
a double peak in the cross-correlation function.

\subsection{No ultra-compact dwarfs in FCSS Fields 2 and 3}
\label{sec_fornaxucdsflds23}

In contrast to the six UCDs in the first of the FCSS 2dF fields, no such 
objects have been found in Fields 2 or 3 despite similar selection 
criteria and survey completeness as Field~1. 
While Field~1 is centered on NGC1399 and the cluster core, these 
two new 2dF fields do still contain numerous Fornax Cluster galaxies, 
both giant and dwarf. We emphasize again that despite surveying out to 
a distance $\simeq 1$~Mpc, all the UCDs detected in the FCSS 
are inside a radius of 
170~kpc of the cluster center. This is strong evidence  
that UCDs are a phenomenon associated with dense environments in 
clusters.

\subsection{Properties of the Fornax ultra-compact dwarfs}
\label{sec_fornaxucdprops}

The new Fornax object increases the sample of bright ($b_j \le 20.0$~mag)
Fornax Cluster UCDs to a total of six. 
All six lie within $28\arcmin$ of NGC1399.

Figure~\ref{fig_raddistn} shows the distribution of 
the Fornax UCDs with angular radius from NGC1399 (the prominent 
central elliptical galaxy of the Fornax Cluster) 
in the form of a cumulative number distribution.
The figure covers the entire radial extent of the 2dF survey, 
extending to 3.0~deg from NGC1399, although the area sampling 
becomes progressively incomplete at larger radial distances 
because of the geometry of the 2dF fields. 
Nevertheless, the figure illustrates clearly the cut-off in the UCDs 
at 0.5~deg, equivalent to a projected angular distance of 170~kpc. 
The distribution of UCDs predicted by Bekki et al.\ (2003) 
using the tidal threshing hypothesis for an age of 8.5~Gyr 
is shown for comparison (see also their Figure~9). 
There is a close match between the shapes of the distributions. 
However, 
Bekki et al.\ predict 14 UCDs formed from the disruption of nucleated 
galaxies, significantly larger in number than the 6 found in the 
FCSS. This is smaller than the 59 faint objects found by Mieske, Hilker \& 
Infante (2004), but their sample will include many conventional 
globular clusters.  
The FCSS sample limits at faint magnitudes mean that fainter 
remnant nuclei of dE,Ns would be overlooked had the UCDs been created 
by the tidal threshing process. Luminosity information for dE,N 
nuclei (e.g.\ Lotz et al.\ 2001) suggests that a majority 
of the Bekki et al.\ predicted UCDs would not get into the 2dF sample 
(given that the threshing model predicts a luminosity function that 
matches the nuclei), 
broadly in line with the observed numbers.

Figure~\ref{fig_raddistn} also shows the distribution of 
nucleated and non-nucleated dwarf elliptical galaxies. 
It should be noted, however, that the numbers of these galaxies 
(90 and 138) are very much larger than the number of UCDs. 
For example, within 0.5~deg of NGC1399, there are 10 dE,Ns and 
13 non-nucleated dEs. 

The mean heliocentric radial velocity of the six bright Fornax UCDs is 
$1460 \pm 109 \:\mbox{km}\mbox{s}^{-1}$
on the basis of the 2dF data, with a standard deviation of 
$266 \pm 78 \:\mbox{km}\mbox{s}^{-1}$ (including the effects of 
observational errors). 
The heliocentric radial velocity of NGC1399 from the NASA Extragalactic 
Database is $1425 \pm 4 \:\mbox{km}\mbox{s}^{-1}$. 
There is therefore no evidence from the 2dF data alone that the 
six bright UCDs have a different systemic velocity to NGC1399. 
Mieske, Hilker \& Infante (2004) reported a slightly higher velocity 
for brighter UCDs than for NGC1399 (and also than for fainter compact 
objects). There is a mean difference of only 
$20 \pm 34 \; \rm km \, s^{-1}$ between the Mieske et al.\  and 
2dF results (for 5 objects in common).

\section{Discussion}
\label{sec_discuss}

None of the 9 Virgo objects shows clear evidence of being resolved
in ground-based CCD imaging or of having extended low surface
brightness halos, but Virgo UCD~4 appears slightly larger than the 
seeing disc in the best seeing exposure available (and incidentally 
has a bluer continuum in its spectrum than the other 8 UCDs). 
Of all 15 bright UCDs across both clusters, only
Fornax UCD~3 is unambiguously resolved in ground-based imaging (which was
discussed in detail by Drinkwater et al.\ 2003).  The failure to find
strong evidence of spatially extended emission around the UCDs 
(beyond $5\arcsec$)
indicates that they are unlikely to be the nuclei of dE,N galaxies
which simply have extremely low surface brightness halos, given the
observed correlation between the surface brightnesses of dE,Ns and
nuclear magnitudes presented in Section~\ref{sec_lsbdens} (see also
Drinkwater et al.\ 2003).  The absence of extended light as faint as
$27-28$~R~mag arcsec$^{-2}$ around seven of the Virgo UCDs also rules
out the possibility that they are nuclei residing in
extremely low surface brightness disks of spiral galaxies.

The available evidence indicates that the bright UCDs are unlikely to
be the high-luminosity tail of the globular cluster system of NGC1399
(a possibility suggested by Drinkwater et al.\ 2000a and Mieske, Hilker \&
Infante 2002).  The bright UCDs in Fornax and Virgo have absolute
magnitudes ($M_B = -10.4$ to $-13.1$) that are brighter than the most
luminous globular cluster in the Galaxy ($\omega$ Centauri, which has
$M_B = -9.5$~mag) and in M31 (G1, with $M_B = -10.3$~mag).  Many
authors, however, have argued that both $\omega$ Cen and G1 might
themselves be the remnant nuclei of tidally disrupted dE,N galaxies
(e.g.\ Meylan et al.\  2001; Bekki \& Freeman 2003; Tsuchiya, Korchagin
\& Dinescu 2004; Bekki \& Chiba 2004).  The size--velocity dispersion
properties of the Fornax UCDs are different to the relations for
globular clusters (Drinkwater et al.\ 2003; Ha\c{s}egan et al. 2005).

The Fornax results show that bright UCDs are spatially concentrated on
the core of the cluster, with the core being defined here as the
bright elliptical galaxy NGC1399.  The UCDs, however, are less
centrally concentrated than the globular cluster system of NGC1399
(Drinkwater et al.\ 2000a), but the difference is more modest when the 
globular cluster distribution of Dirsch et al.\  (2003) is used in preference 
to earlier data. The radial distribution of Virgo UCDs
around M87 (Section~\ref{sec_virgoradialdistn}) is much more 
extended than that of the M87 globular cluster system revealed in 
published studies, but an accurate comparison is difficult because 
the available data on the globular clusters are restricted to the 
immediate vicinity of M87. 
There is no convincing evidence (Section~\ref{sec_radvels}) that the 
mean radial velocity of the Virgo UCDs, or their velocity dispersion, 
is different to that of M87 or the M87 globular cluster system. 
Such a result which if found would indicate that the UCDs are not 
simply a more luminous component of the M87 globular cluster system, 
but a free-floating system in the core of the Virgo Cluster.  Bekki et
al.\ (2003) showed that dE,N galaxies are more likely to be tidally
stripped if their orbits within the cluster have high eccentricities,
a process that would produce a system of UCDs with a larger velocity
dispersion than the system of dE,Ns under the tidal threshing model.
The velocity dispersion of the 9 UCDs, however, is slightly smaller
than that of dE,N galaxies in the same area satisfying the same
velocity selection criteria, although the two values are
statistically consistent.

\begin{figure}
\hspace{2mm}
\includegraphics[width=70mm,angle=0.]{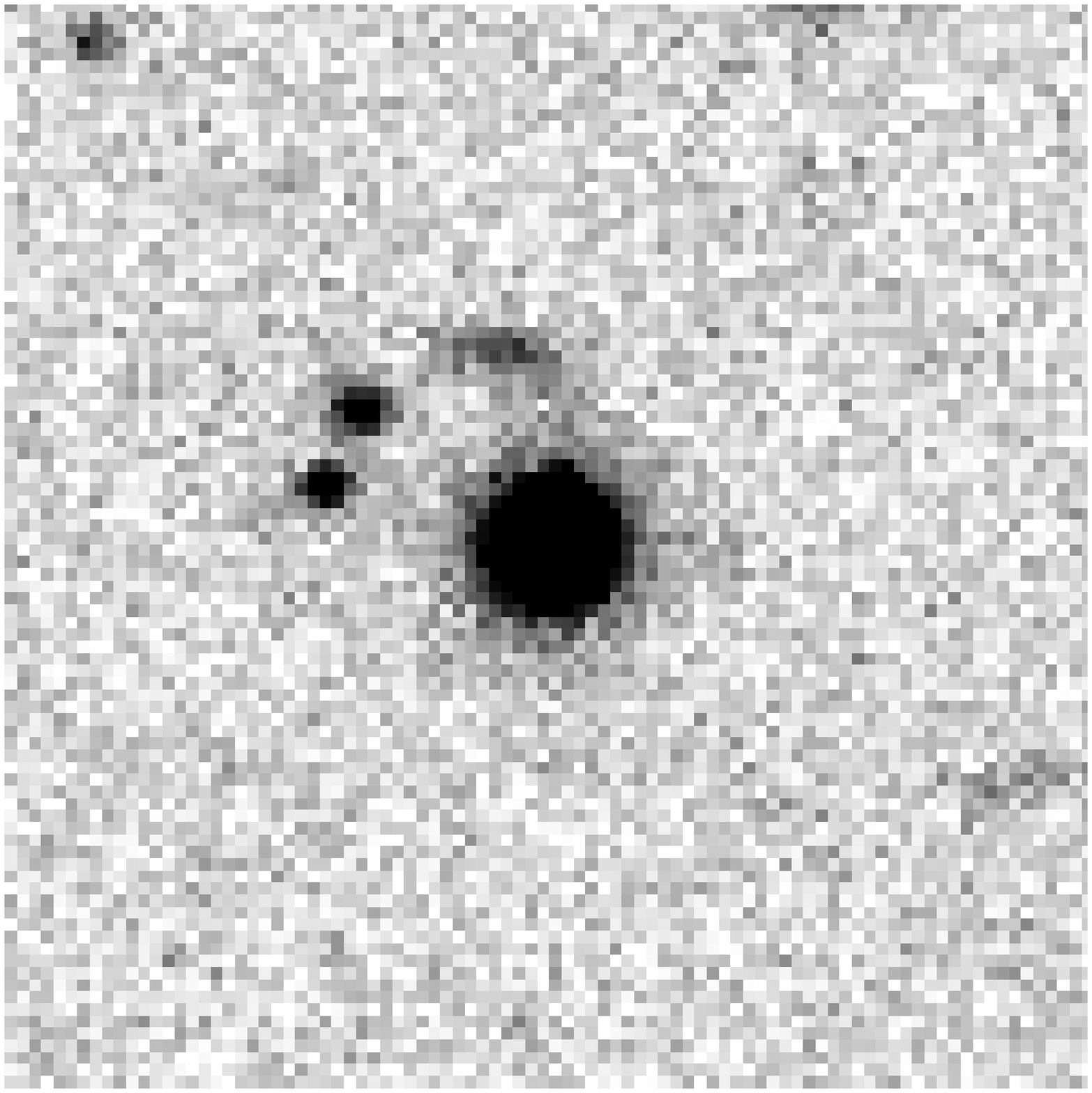}
\caption{
A CCD image of Virgo UCD 6.  A $30\arcsec$ $\times$ $30\arcsec$ 
region around UCD 6 is shown based on a 25~min B-band integration 
with the Isaac Newton Telescope in $1\farcs1$ seeing. 
}
\label{fig_intwfs_vucda6}
\end{figure}

Fornax UCD~3 consists of a bright, highly compact core within a more
extended envelope. It is difficult to explain such an object as being
an unusually luminous globular cluster, whereas objects of this kind
are explicitly predicted by the tidal threshing model (Bekki et
al.\ 2003).  The tidal threshing model also predicts that some UCDs may
be accompanied by discrete fainter companion objects, the result of
tidal debris or globular clusters from the progenitor galaxy.  One of
the Virgo objects, UCD~6, has three objects lying nearby, as discussed
in Sections~\ref{sec_sbprofiles} and~\ref{sec_lsbhalolimits}. 
The nearby images 
might be expected to have colours similar to the UCD in the threshing 
model were they physically associated. 
The result from Section~\ref{sec_sbprofiles} that the colours are 
different suggests that they may be background galaxies/foreground stars.

Ha\c{s}egan et al. (2005) performed a careful study of objects in 
the immediate vicinity of M87 to investigate the connection  
between UCDs and the most luminous M87 globular clusters. 
Of the six objects around M87 for which they had  high-resolution 
spectroscopy and HST imaging, two had properties indicating that 
they are likely to by very luminous globular clusters. The other 
four -- on the basis of their sizes, velocity dispersions, 
colours and mass-to-light ratios -- had properties more consistent 
with UCDs. Significantly, the mass-to-light ratios are higher 
than can be expected on the basis of normal stellar populations 
alone, indicating the likely presence of dark matter. In contrast, 
Drinkwater et al. (2003) found mass-to-light ratios for Fornax 
UCDs that, although high, did not unambiguously indicate the 
presence of dark matter. 
Ha\c{s}egan et al. noted that one of their four M87 UCDs has 
properties consistent with the Fellhauer \& Kroupa (2002) 
model for producing UCDs from star clusters formed in galaxy 
interactions. 

Ha\c{s}egan et al. (2005) also identified candidate UCDs from HST 
Advanced Camera for Surveys imaging (C\^{o}t\'{e} et al.\  2004), 
using either published velocity data or surface brightness 
fluctuations to assess the likelihood that they lie in the Virgo Cluster. 
Some of these show evidence of small haloes, like Fornax UCD3 
of Drinkwater et al.\  (2000a; see also Hilker et al. 1999, 
Phillipps et al.\  2001), but in contrast to the nine objects 
described in Section~\ref{sec_virgo}. 
Such compact haloes are explicitly predicted by the tidal threshing 
model of Bekki, Couch, \& Drinkwater (2001).

De Propris et al. (2005) found that the nuclei of a sample of 
18 Virgo Cluster dE,N galaxies drawn from HST archive observations 
are generally smaller than the 5 Fornax UCDs of Drinkwater et al.\  
(2000a). This could be interpreted as evidence against the tidal 
threshing model. However, a direct comparison between the two 
types of object is not straightforward because the five UCDs have 
absolute magnitudes that are generally brighter than the dE,N nuclei, 
having been identified in a survey of targets selected by apparent 
magnitude. Further observations of the most luminous dE,N nuclei 
will be necessary to clarify the connection between the brightest 
UCDs and nuclei.

The results presented here, including the comparison between the
observed and predicted radial distributions within the Fornax Cluster,
are broadly consistent with the predictions of the tidal threshing 
model (Bekki et al.\
2001; Bekki et al.\ 2003) (Section~\ref{sec_fornaxucdprops}).  
However, the Bekki et al.\ (2003) predicted UCD radial distribution 
around M87 extends as far as $2\fdg5$, whereas the 2dF observations 
here extend only $1\fdg0$ from M87: a direct comparison between 
the observed and predicted UCDs distributions is not possible in 
the Virgo Cluster. 

The Virgo compact elliptical galaxy NGC4486B, one of the few galaxies with
properties comparable to M32, lies in the region of M87, between M87
and Virgo UCD1 in Figure~\ref{fig_virgoucds}.  Tidal stripping of its
outer regions by M87 has been advocated as the cause of NGC4486B's
unusual morphology (e.g.\ Faber 1973).
The distribution of the UCDs does not appear to be symmetric about M87: 
8 of the 9 are found northeast of a line running southeast--northwest 
through M87 (which corresponds approximately to the major axis of M87).

The new discoveries of fainter compact objects in the Fornax Cluster
by Mieske, Hilker \& Infante (2004) and Drinkwater et al.\ (2004, 
and in preparation) suggest that the bright UCDs (described in this paper,
and by Hilker et al.\ 1999, Drinkwater et al.\ 2000a and Phillipps et
al.\ 2001) may be only the bright tail of a population of compact
objects in cluster cores.  The luminosity function of the nuclei of
dE,N galaxies  (Lotz et al.\ 2001)  spans the luminosities of the 
bright UCDs and extends to lower luminosities. The tidal threshing model
therefore predicts a population of fainter compact objects in the
cores of galaxy clusters.  Bekki et al.\ (2003) predict 14 UCDs of all
magnitudes in Fornax and 46 in Virgo.  The predicted number in Fornax
is significantly smaller than the numbers of faint compact objects of
Mieske, Hilker \& Infante (2004) and Drinkwater et al.\ (in
preparation).  The actual numbers of compact objects in the cluster
cores expected from the tidal threshing model will be larger than the
numbers of naked nuclei themselves because of the presence of globular
clusters released from tidally disrupted galaxies (although these free globular
clusters will have significantly lower luminosities than the naked
nuclei), including tidally disrupted luminous galaxies as well as 
dwarfs  (see Bassino et al. 2003 for a discussion of intracluster 
globular clusters). 
The discovery of these objects in the nearest galaxy clusters is 
complemented by the identification of candidate UCDs in the 
Abell 1689 cluster (at a redshift $z = 0.18$) from imaging data by 
Mieske et al.\ (2004).

Mieske, Hilker \& Infante (2004) compared the colors of Fornax UCDs
with globular clusters and dE,N nuclei.  The bright UCDs had
(V--I)$_J$ colors that were slightly redder by 0.1~mag than the
nuclei of dE,N galaxies.  The tidal threshing hypothesis would predict
that the nuclei and UCDs would have similar colors if star formation
in the nuclei in dE,Ns had ceased before the disruption processes
which produced the UCDs had begun.  Had star formation continued in
the nuclei of the surviving dE,Ns after tidal threshing had begun in
those galaxies destined to become UCDs, the surviving dE,Ns could have
bluer colors than the UCDs today.  The observation that UCDs are
redder than dE,N nuclei could therefore be interpreted as evidence
that tidal threshing curtailed star formation in dE,N nuclei before
nucleus star formation was complete.

Fellhauer \& Kroupa (2002) argued that UCDs might be the result of
mergers between massive star clusters formed in galaxy interactions.
These merged objects will age following the end of their star
formation to produce compact objects with early-type galaxy spectra and
half-light radii $\simeq 40$ to 160~pc. These are comparable to, or
larger than, the observed sizes of the first five Fornax bright UCDs
determined with the HST (Drinkwater et al.\ 2003).  Maraston et
al.\ (2004) measured the velocity dispersion of the stellar population
in the giant star cluster W3 in the merged galaxy NGC7252, finding a
value of $45 \pm 5$~km$\,$s$^{-1}$.  This is larger by factors ranging
from 1.2 to 1.9 than the first five Fornax UCDs (Drinkwater et
al.\ 2003).  W3 is likely to evolve over time into an object similar to
the bright UCDs.  Massive star clusters can therefore explain many of
the observed properties of UCDs.  Fellhauer \& Kroupa (2005) modelled 
the formation of W3 by the merging of star clusters formed in the 
galaxy interaction which produced NGC7252. They predict that W3 has 
a compact core surrounded by a more extended halo; this has similarities 
to Fornax UCD3, which has a halo with an exponential scale length 
of 60~pc. UCDs with such envelopes are also directly predicted by 
the tidal threshing model.

Further observations in the Virgo Cluster will be needed to establish
the distance from M87 at which the UCD radial distribution falls off,
which would be an important test of the models for the origin of the
UCDs.  Bekki et al.\ (2003) predicted a cutoff at $2\fdg5$ (700~kpc)
from M87 in the tidal threshing model, significantly beyond the limits
of the current survey (which currently reaches $1\degr$, with poorer
sampling beyond $0\fdg6$ from M87).

\section{Conclusions}
\label{sec_concls}

New 2dF observations of star-like objects in a $1\fdg0$ radius region
centered on M87 in the Virgo Cluster have found 9 objects with
properties similar to the bright Fornax UCDs.  The discovery of UCD
galaxies in the Virgo Cluster shows that this recently recognized
class of objects is not peculiar to Fornax, and implies that UCDs are
a common phenomenon in all galaxy clusters.  These objects have sizes
$\ll 100$~pc and blue absolute magnitudes $M_B = -10$ to $-13$; they
are not clearly resolved by ground-based CCD imaging and do not have 
extended low surface brightness halos.  If they can be found in sufficient
numbers, the UCDs would be important test particles riding in the
cluster gravitational potential.

We describe the properties of a sixth bright UCD in the Fornax Cluster 
using data from the 2dF Fornax Cluster Spectroscopic Survey. 
It is unresolved in available
Schmidt photographic imaging data, has a blue photographic magnitude
of $b_j = 19.4$ (equivalent to a blue absolute magnitude of $M_B =
-12.0$), and has an early-type galaxy spectrum.  It is more luminous
than most of the objects in the recently discovered fainter compact-object
population of Mieske, Hilker \& Infante (2004) and Drinkwater et al.\
(in preparation).  With new data from two additional fields, the Fornax 2dF
observations extend to $3\degr$ (equivalent to a projected distance of
1.0~Mpc) from NGC1399, the bright elliptical galaxy at the core of the
cluster.  The observations have established that within the  Fornax
Cluster, bright UCDs lie exclusively within $0\fdg5$ of NGC1399.  The
cutoff in their radial distribution (a projected distance of 170~kpc
from NGC1399) is very close to that predicted by Bekki et al.\ (2001)
in the tidal threshing model in which the UCDs are the remnant nuclei
of galaxies that suffered tidal disruption due to repeated passes of a
massive galaxy (NGC1399 in the case of the Fornax Cluster).  Their 
distribution is slightly broader than that of the NGC1399 globular 
cluster system. 

The available evidence is consistent with the UCDs being the remnant
nuclei liberated via tidal threshing of dE,N galaxies.  Many of the 
observed properties are also consistent with models of luminous 
star clusters formed in galaxy interactions.   Seen in this
light, characterizing the properties of the UCD population in a galaxy
cluster is a potential gauge of the long-term effects of
tidally-induced evolution over the lifetime of a galaxy cluster.

\bigskip

\section*{Acknowledgements}

These observations were made using the 2dF spectrograph at the 
Anglo-Australian Observatory. We wish to thank the support 
and encouragement given by the AAO staff. This work made 
use of data from the United Kingdom Schmidt Telescope scanned by 
the APM Machine. 
Other imaging data were recorded with the Isaac Newton Telescope 
of the Isaac Newton Group in La Palma and provided by the 
ING Archive at the U.K. Astronomy Data Centre, Cambridge. 
Other photographic data from the United Kingdom Schmidt Telescope 
were scanned by the SuperCOSMOS plate machine at the Royal Observatory 
Edinburgh. The stack of 63 UKST films was produced by the 
SuperCOSMOS Unit at the Royal Observatory Edinburgh. 

Part of the work reported here was done at the Institute of Geophysics
and Planetary Physics, under the auspices of the U.S. Department of
Energy by Lawrence Livermore National Laboratory under contract
No.~W-7405-Eng-48 and has also been supported by the National Science
Foundation under grant No.~0407445.

We thank the anonymous referee for detailed comments that have 
improved the paper.

\end{document}